\documentclass[11pt,a4paper]{article}
\pdfoutput=1
\usepackage{jheppub}

\usepackage{multirow, graphicx,amssymb,url,mathrsfs,amsmath}
\usepackage{wrapfig,boxedminipage,setspace,subfigure,epsfig}
\usepackage{amsxtra,amstext,latexsym,dsfont,amsfonts}
\usepackage{color,eucal}
\usepackage[dvipsnames]{xcolor}
\usepackage{float}
\usepackage{slashed,comment}
\usepackage{kotex}

\usepackage{tikz}
\usetikzlibrary{calc,patterns,angles,quotes}
\usetikzlibrary{decorations.pathreplacing,decorations.markings,snakes}

\usepackage{tabularx, array}
\newcolumntype{L}[1]{>{\raggedright\arraybackslash}p{#1}}
\newcolumntype{C}[1]{>{\centering\arraybackslash}p{#1}}
\newcolumntype{R}[1]{>{\raggedleft\arraybackslash}p{#1}}

\newcommand{\dd}{\mathrm{d}}

\title{Holographic entanglement density for spontaneous symmetry breaking}

\author[a,b]{Hyun-Sik Jeong,}
\author[c,d]{Keun-Young Kim,}
\author[a,b]{and Ya-Wen Sun}

\emailAdd{hyunsik@ucas.ac.cn}
\emailAdd{fortoe@gist.ac.kr}
\emailAdd{yawen.sun@ucas.ac.cn}

\affiliation[a]{School of physics $\&$ CAS Center for Excellence in Topological Quantum Computation, University of Chinese Academy of Sciences, Zhongguancun east road 80, Beijing 100049, China}
\affiliation[b]{Kavli Institute for Theoretical Sciences, University of Chinese Academy of Sciences, \\ Zhongguancun east road 80, Beijing 100049, China}
\affiliation[c]{Department of Physics and Photon Science, Gwangju Institute of Science and Technology, \\
123 Cheomdan-gwagiro, Gwangju 61005, Korea}
\affiliation[d]{Research Center for Photon Science Technology, Gwangju Institute of Science and Technology, \\
123 Cheomdan-gwagiro, Gwangju 61005, Korea}

\abstract{
We investigate the properties of the holographic entanglement entropy of the systems in which the $U(1)$ or the translational symmetry is broken \textit{spontaneously}. For this purpose, we define the entanglement density of the strip-subsystems and examine both the first law of entanglement entropy (FLEE) and the area theorem. We classify the conditions that FLEE and/or the area theorem obey and show that such a classification may be useful for characterizing the systems. We also find universalities from both FLEE and the area theorem. In the spontaneous symmetry breaking case, FLEE is always obeyed regardless of the type of symmetry: $U(1)$ or translation. For the translational symmetry, the area theorem is always violated when the symmetry is weakly broken, independent of the symmetry breaking patterns (explicit or spontaneous). 
We also argue that the $\log$ contribution of the entanglement entropy from the Goldstone mode may not appear in the strongly coupled systems.
}

\begin{document}
\maketitle

%
\section{Introduction}

Entanglement entropy has been playing an important role in the investigation of quantum gravity and quantum field theory. In particular, with the Ryu-Takayanagi formula~\cite{Ryu:2006bv,Ryu:2006ef}, the holographic duality (or gauge/gravity duality)~\cite{Maldacena:1997re,Witten:1998qj,Gubser:1998bc} provides a remarkable connection between gravity theory and  conformal field theory: a geometric quantity in the bulk spacetime can be related to the entanglement entropy of the boundary field theory. In other words, the Ryu-Takayanagi formula gives us a hint to understand the emergence of spacetime from the entanglement properties in the dual field theory (e.g., see also \cite{Nishioka:2009un,VanRaamsdonk:2010pw,Nozaki:2012zj,Lin:2014hva,Hayden:2016cfa}).

In addition to its role in quantum gravity, entanglement entropy has also been used in condensed matter physics: entanglement entropy may also play an important role in characterizing and classifying the phases of matter.\footnote{See \cite{Laflorencie:2015eck} for a review of quantum entanglement in condensed matter physics. See also \cite{Zhang:2022yaw} for recent progress in quantum entanglement in many-body systems using SYK model and its generalizations.}
For instance, entanglement entropy between the subsystem and the rest of the system could exhibit characteristic behavior of entanglement entropy with the subsystem size $\ell$: for ($d+1$) dimensions,  entanglement entropy has a $\log (\ell^{d})$ contribution from the Goldstone boson~\cite{Metlitski:2011pr}, $\ell^{d} \log \ell$ from the Fermi surface~\cite{Wolf:2006zzb,Gioev:2006zz,Swingle:2009bf,Swingle:2010yi}, or $\ell$-independent behavior from the topologically ordered degrees of freedom~\cite{Kitaev:2005dm,Levin:2006zz,Castelnovo_2008,Grover:2011fa}.

\paragraph{Entanglement entropy in holography:}
Inspired from the properties of entanglement entropy with the subsystem size $\ell$ in condensed matter physics, it is instructive to investigate the $\ell$-dependence of entanglement entropy in holography.

Using the Ryu-Takayanagi formula \eqref{HEEFOR}, one can find interesting features of entanglement entropy in the small ($\ell\ll1$) or large ($\ell\gg1$) subsystem limit.
For the \textit{small} subsystem, ``entanglement thermodynamics" may apply. In particular, when the system is excited entanglement entropy at $\ell\ll1$ may obey a property analogous to the first law of thermodynamics, called \textbf{\textit{the first law of entanglement entropy|}} (FLEE)~\cite{Bhattacharya:2012mi}\footnote{FLEE has been extensively investigated in many holographic models~\cite{Guo:2013aca,Allahbakhshi:2013rda,He:2013rsa,Park:2015hcz,Lin:2017svk,Ghosh:2016fop,Sun:2016dch,OBannon:2016exv,Bhattacharya:2017gzt,Bhattacharya:2019zkb,Lokhande:2017jik,Caceres:2016xjz,Blanco:2013joa,Lin:2014hva,He:2014lfa,Dong:2013qoa,Pal:2015mda,Sun:2016til,Bueno:2016gnv,Haehl:2017sot,Gushterov:2017vnr,Erdmenger:2017pfh,Nadi:2019bqu,Saha:2019ado,Fujita:2020qvp,Maulik:2020tzm,Santos:2022oyo}.}:
\begin{align}\label{FLEEEQF}
\begin{split}
\Delta S = \frac{1}{T_{\text{ent}}} \, \Delta E \,, \qquad (\ell\ll1)
\end{split}
\end{align}
where $\Delta S:=S-S_{\text{CFT}}$ is the increased amount of the entanglement entropy in excited states, $S$, compared with the ground state of the CFT, $S_{\text{CFT}}$ \eqref{HEEFORFIN3}. $\Delta E$ is the corresponding amount of energy in the subsystem given by
\begin{equation}\label{DELTAENF}
\Delta E = \int \dd^d x \, \langle T_{tt} \rangle \,,
\end{equation}
where $\langle T_{tt} \rangle$ is the energy density of the excited state.\footnote{Note that $\langle T_{tt} \rangle$ of the ground state of the CFT (or the pure AdS) is zero so that $\Delta \langle T_{tt} \rangle = \langle T_{tt} \rangle$.} 
$T_{\text{ent}}$ in \eqref{FLEEEQF} is called the entanglement temperature~\cite{Bhattacharya:2012mi,Blanco:2013joa}, which is proportional to the inverse of the subsystem size as $T_{\text{ent}}\sim1/\ell$. Note that $T_{\text{ent}}$ is universal in that it only depends on the shape of the subsystem and the dimension $d$, i.e., $T_{\text{ent}}$ may not depend on the details of the excited states.

In this paper, we consider a strip subsystem of width $\ell$ in the $x$-direction in $d=2$ (i.e., the dual gravity is  asymptotically AdS$_{4}$): see Fig. \ref{STRIPfig}. In this setup \cite{Bhattacharya:2012mi}, $\Delta E$ in \eqref{DELTAENF} and $T_{\text{ent}}$ is
\begin{equation}\label{TENT}
\Delta E = \ell \, \Omega \,  \langle T_{tt} \rangle\,, \quad T_{\text{ent}} = \left(\frac{4\,\Gamma\left(\frac{3}{4}\right)}{\Gamma\left(\frac{1}{4}\right)}\right)^{2} \frac{1}{\pi \, \ell} \,,
\end{equation}
where $\Omega$ is the length in the $y$-direction.
Thus, FLEE \eqref{FLEEEQF} implies $\Delta S \sim \langle T_{tt} \rangle \, \ell^2$ at small $\ell$.

For the \textit{large} subsystem, the entanglement entropy of the excited state, $S$, behaves as     
\begin{align}\label{ATVQF}
\begin{split}
S = s \, V \,+\,  \alpha \, A  \,+\,  \dots   \,, \qquad (\ell\gg1)
\end{split}
\end{align}
where $s$ is the thermal entropy density, $\alpha$ is a dimensionful constant, and $\dots$ denotes further sub-leading terms of $\mathcal{O}(\ell^{-1})$.
Note that \eqref{ATVQF} consists of two terms: i) the ``volume law" term ($\propto V= \ell \, \Omega$) in which $V$ is the volume of the strip; ii) the ``area law" term ($\propto A= 2 \, \Omega$) where $A$ is the area of the one-dimensional boundary of the strip.\footnote{\eqref{ATVQF} may also contain other types of area terms such as $A \log A$ which corresponds to an area law violation~\cite{Swingle:2011np}}. 

The volume law term is the leading contribution of $S$, which may be expected for excited states, e.g., a thermal state, in that when $\ell\rightarrow\infty$ the subsystem becomes the entire system so that the minimal surface lies along the horizon~\cite{Hubeny:2012ry,Liu:2013una} implying that the Ryu-Takayanagi formula \eqref{HEEFOR} may be related with the thermal entropy density $s$.

The area law term~\cite{Bombelli:1986rw,Srednicki:1993im,Hastings:2007iok,Eisert:2008ur} is a sub-leading contribution of $S$ with the parameter $\alpha$ in which $\alpha$ could be expressed as a sum of several terms, in general.
For instance, for the vacuum state of the CFT (in which $s=0$), $S_{\text{CFT}}$ in \eqref{HEEFORFIN3}, $\alpha$ is a sum of two terms: one from the UV divergence $1/\epsilon$, the other $1/\ell$~\cite{Ryu:2006bv,Ryu:2006ef}.

In addition to the appearance of thermal entropy density ($s$) in the leading term of $S$, one can also find an interesting property from the sub-leading term with $\alpha$: \textbf{\textit{the area theorem}}~\cite{Ryu:2006ef,Myers:2012ed,Casini:2012ei,Casini:2016udt}. The area theorem is a variant of the $c$-theorem\footnote{Considering how entanglement entropy behaves under a renormalization group (RG) flow, the interesting theorem, the $c$-theorem~\cite{Zamolodchikov:1986gt}, has been derived in the two-dimensional conformal field theory, which states that the central charge $c$ decreases along the RG flow as
\begin{equation}\label{CTOM}
c_{\text{UV}} \geq c_{\text{IR}} \,,
\end{equation}
where the $c$-function is a monotonically decreasing function of the energy scale and equals the central charge $c$ at fixed points. For higher dimensions, there are other type of theorems without using the central charge (recall that  there are more than two central charges in higher dimension). For instance, the $a$-theorem~\cite{Cardy:1988cwa,Komargodski:2011vj,Komargodski:2011xv} with the anomaly $a$ and the $F$-theorem~\cite{Jafferis:2011zi,Klebanov:2011gs} with the free energy $F$. See also \cite{Myers:2010xs,Myers:2010tj,Casini:2012ei,Casini:2004bw,Chu:2019uoh,Park:2018ebm,Park:2019pzo,Giataganas:2017koz,Baggioli:2020cld,Cremonini:2020rdx,Hoyos:2021vhl,Cartwright:2021hpv,Arefeva:2020uec,Cremonini:2013ipa}.}, which states that the value of $\alpha$ at UV/IR fixed points obeys the following inequality
\begin{equation}\label{ATOM}
\alpha_{\text{UV}} \geq \alpha_{\text{IR}} \,,
\end{equation}
where the field theoretic proof was given for a sphere in $d=3$ with strong sub-additivity in \cite{Casini:2012ei} or for a sphere in $d\geq3$ with the positivity of relative entropy~\cite{Casini:2016udt}, and for a strip in $d\geq3$ with the Null Energy Condition (NEC)~\cite{Myers:2012ed} in which the NEC may be holographically dual to strong sub-additivity~\cite{Headrick:2007km,Wall:2012uf}.

\paragraph{Entanglement density $\sigma$:}
In order to examine $\alpha$, it is useful to introduce the entanglement density $\sigma$~\cite{Gushterov:2017vnr}\footnote{$\sigma$ is not the entanglement density~\cite{Nozaki:2013wia,Bhattacharya:2014vja} defined as the second derivative of entanglement entropy.} defined as 
\begin{equation}\label{SIGFORF}
\sigma  :=  \frac{S - S_{\text{CFT}}}{V} =  \frac{\Delta S}{V} \,,
\end{equation}
which yields a finite value because the UV divergence $1/\epsilon$ is canceled out.
In terms of $\sigma$ \eqref{SIGFORF}, the large subsystem behavior of the entanglement entropy \eqref{ATVQF} can be expressed as 
\begin{align}\label{ATVQF2}
\begin{split}
\sigma = s \,-\,  \Delta \alpha \, \frac{A}{V}  \,+\,  \dots   \,, \qquad (\ell\gg1)
\end{split}
\end{align}
where the sub-leading term, $A/V$, is order of $\mathcal{O}(1/\ell)$ and $\Delta \alpha$ is defined 
\begin{align}\label{withareathemalt}
\begin{split}
\Delta \alpha = \alpha_{\text{CFT}}-\alpha \,,
\end{split}
\end{align}
which may be interpreted as  $\alpha_{\text{UV}}-\alpha_{\text{IR}}$~\cite{Gushterov:2017vnr,Erdmenger:2017pfh,Giataganas:2021jbj}.\footnote{Recall that $s=0$ for the entanglement entropy of the pure AdS geometry ($S_{\text{CFT}}$).}
Thus, using \eqref{withareathemalt}, the area theorem \eqref{ATOM} can be rephrased as $\Delta \alpha\geq0$.
%

\begin{figure}[]
\centering
     \includegraphics[width=10.0cm]{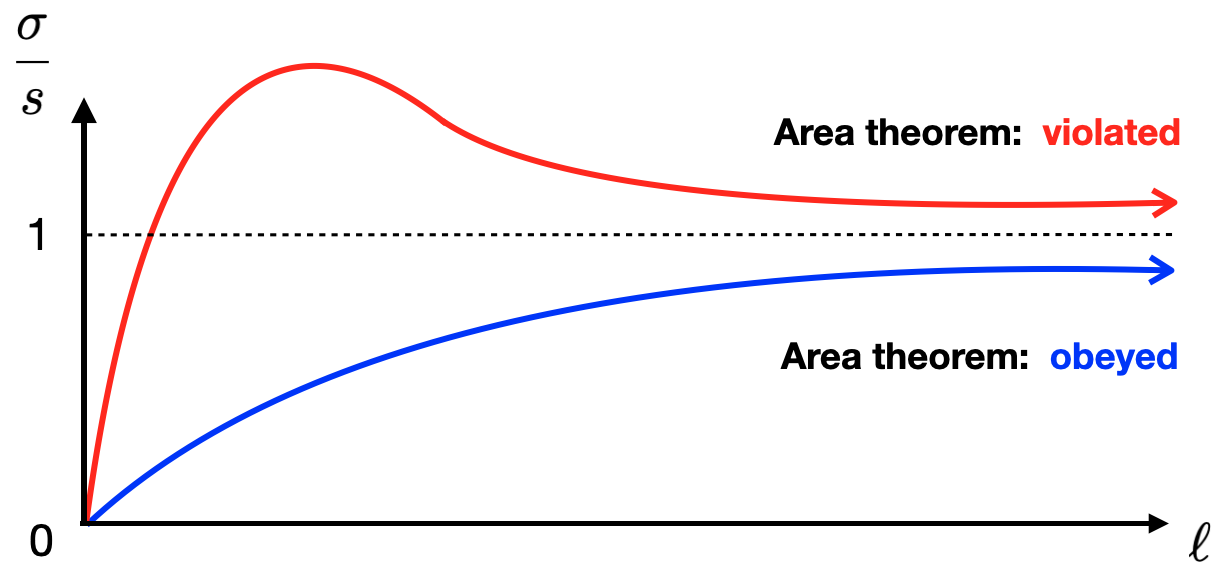}
\caption{A schematically figure for $\sigma$ in unit of $s$. The first law of entanglement entropy (FLEE) implies $\sigma/s\sim\ell$ at small $\ell$. In the large $\ell$ regime, the area theorem can be obeyed when $\sigma/s\rightarrow1^{-}$ (blue line) or violated when $\sigma/s\rightarrow1^{+}$ (red line). }\label{cartoonfig}
\end{figure}

\paragraph{Motivations of this paper :}
In summary, using $\sigma$ \eqref{SIGFORF}, the $\ell$-dependence of holographic entanglement entropy in the small/large subsystem limit can be expressed as 
\begin{align}\label{SIGFORFfin}
\begin{split}
(\text{The first law of entanglement entropy}):  \quad &\sigma = \langle T_{tt} \rangle \, T_{\text{ent}}^{-1} \,, \qquad\qquad\qquad\qquad\quad \,\,\, (\ell \ll 1) \\
(\text{The area theorem}):  \quad &\sigma = s \,-\,  \Delta \alpha \, \frac{A}{V}  \quad\,\,\, \text{with} \quad \Delta \alpha \geq 0  \,, \quad (\ell \gg 1) 
\end{split}
\end{align}
in which $T_{\text{ent}}^{-1}\sim\ell$ and $\frac{A}{V}\sim\ell^{-1}$.\footnote{Note that $\Delta E$ in \eqref{TENT} is  $V \,  \langle T_{tt} \rangle$. Thus, $\Delta S$ in \eqref{FLEEEQF} is $V \,  \langle T_{tt} \rangle \, T_{\text{ent}}^{-1}$ so that $\sigma = \langle T_{tt} \rangle \, T_{\text{ent}}^{-1}$.}
Note that although $\sigma$ basically has the same information with $\Delta S$ because dividing by $V$ is technically trivial, $\sigma$ is a practically useful quantity in that we can easily check the area theorem $\Delta \alpha \geq0$ by the naked eye: if $\sigma\rightarrow s^{-}$ at $\ell\gg1$, this implies $\Delta \alpha \geq0$. See Fig. \ref{cartoonfig}.

Using \eqref{SIGFORFfin}, it is proposed \cite{Gushterov:2017vnr} that $\sigma$ may be used to characterize and classify the states of matter (or geometries in holography).
In particular, depending on if FLEE and/or the area theorem is violated or not, various AdS black holes have been classified in \cite{Gushterov:2017vnr}: the AdS domain wall, the hyperscaling-violating black hole, the AdS soliton, the neutral black hole, the charged black hole, the black hole with a broken translational symmetry.\footnote{See also \cite{Erdmenger:2017pfh} for the analysis of $\sigma$ with the neutral black hole, \cite{Giataganas:2021jbj} for $\sigma$ in a large dimension limit.}

In this paper, using \eqref{SIGFORFfin} we apply the analysis given in \cite{Gushterov:2017vnr} to the case in which  symmetry is broken \textit{spontaneously}. 
In particular, we consider the most well-studied symmetries for the strongly correlated systems in holography~\cite{Hartnoll:2016apf,Zaanen:2015oix,Ammon:2015wua,Baggioli:2019rrs}: the $U(1)$ symmetry and the translational symmetry which can have applications to superconductors and charge density waves, respectively.

Our goal is to examine if the obedience/violation of FLEE/area theorem in \eqref{SIGFORFfin} can also be used to classify the phases with \textit{spontaneous} symmetry breaking, i.e., we extend the analysis in \cite{Gushterov:2017vnr} to more realistic condensed matter phases in holography.
Also, our work can be complementary to the case with a broken translational symmetry in \cite{Gushterov:2017vnr} in that the translational symmetry was broken \textit{explicitly} in \cite{Gushterov:2017vnr}, while it is broken \textit{spontaneously} in our paper.

Note also that entanglement entropy for the superconducting phase (the $U(1)$ symmetry breaking) has been extensively investigated in holography~\cite{Albash:2012pd,Takayanagi:2012kg,Cai:2012sk,Cai:2012es,Cai:2012nm,Cai:2013oma,Li:2013rhw,Johnson:2013dka,Dutta:2013osl,Kuang:2014kha,Peng:2014ira,Garcia-Garcia:2015emb,Peng:2015yaa,Peng:2015mzj,Liu:2015lit,Yao:2016ils,Zeng:2016fsb,Peng:2016jor,Zangeneh:2017tub,Das:2017gjy,Yao:2018shm,Dudal:2018ztm,Guo:2019vni,LalehganiDezaki:2019ykk,Liu:2020blk,Chen:2021vzm,Yao:2021cns}.
However, to our knowledge, our approach (FLEE/area theorem) with \eqref{SIGFORFfin} has not been investigated yet for holographic superconductors. 

Moreover, even for the translational symmetry breaking, entanglement entropy has only been studied with the explicit breaking case~\cite{RezaMohammadiMozaffar:2016lbo,Tanhayi:2016uui,OBannon:2016exv,Gushterov:2017vnr,Kim:2018mgz,Li:2019rpp,Huang:2019zph,Liu:2021rks,Cheng:2021hbw} and the analysis with the spontaneous breaking is still missing. Thus, in this paper, the entanglement entropy with spontaneously broken translational symmetry is analyzed for the first time.

It may also be important to make a further comment that the precise condition when the area theorem is violated ($\Delta \alpha <0$) has not been fully understood yet. From the low temperature analysis with the various black hole geometries in \cite{Gushterov:2017vnr}, it is argued that the near-horizon geometry may be related to the area theorem violation: e.g., the black hole with the AdS$_2\times R^2$ IR geometry could violate the area theorem. Thus, using spontaneous symmetry breaking, we make one step further in this direction and attempt to have a more complete understanding of the area theorem.

This paper is organized as follows.
In section \ref{sec2label}, we review entanglement entropy in holography and also introduce the useful quantity, $\sigma$, to examine \eqref{SIGFORFfin}.
In section \ref{sec3label}, using the holographic superconductor model, we study the obedience/violation of the FLEE/area theorem in which the $U(1)$ symmetry is broken.
In section \ref{sec4label}, we study how the (explicitly or spontaneously) broken translational symmetry can affect both FLEE and the area theorem with the holographic axion model.
Section \ref{sec5label} is devoted to conclusions.

%
\section{Setup}\label{sec2label}

In this section, we will review the holographic entanglement entropy in the asymptotically AdS$_4$ metric \eqref{METRIC} and also introduce the entanglement density $\sigma$ \eqref{SIGFORF} in terms of the function of the metric. 
Also, we further express $\sigma$ in units of the thermal entropy density $s$, $\sigma/s$, in order to make $\sigma$ dimensionless. Such a quantity is useful not only because it is dimensionless, but also for the convenient evaluation of \eqref{SIGFORFfin} in numerics (in particular for the area theorem, e.g., see Fig. \ref{cartoonfig}).

Note that our metric \eqref{METRIC} becomes the one in \cite{Gushterov:2017vnr} when $h(z)=1$. Thus, our analysis in this section corresponds to the generalization of \cite{Gushterov:2017vnr} to the case with more general metrics.

\subsection{Holographic entanglement density}
We consider the asymptotically AdS$_4$ metric:
\begin{equation}\label{METRIC}
\dd s^2 = \frac{L^{2}}{z^2} \left[-f(z) \dd t^2 + \frac{\dd z^2}{g(z)} + h(z) (\dd x^2 + \dd y^2) \right] \,,
\end{equation}
with an AdS radius $L$. The functions $f(z)$, $g(z)$ and $h(z)$ in \eqref{METRIC} are expanded near the AdS boundary ($z\rightarrow0$) as
\begin{equation}\label{ADSASMPF}
f(z) = 1 - \sum_{i=1} \, \mathbf{f}_{i} \, z^{i}  \,, \quad g(z) = 1 - \sum_{i=1} \, \mathbf{g}_{i} \, z^{i} \,, \quad h(z) = 1 - \sum_{i=1} \, \mathbf{h}_{i} \, z^{i} \,,
\end{equation}
where $\mathbf{f}_{i}$, $\mathbf{g}_{i}$, ang $\mathbf{h}_{i}$ are model dependent constants.
From the holographic renormalization~\cite{Balasubramanian:1999re,deHaro:2000vlm}, the asymptotic of the metric \eqref{ADSASMPF} determines the energy density of the dual field theory, $\langle T_{tt} \rangle$, as
\begin{equation}\label{EDFTF}
\langle T_{tt} \rangle = \frac{L^{2}}{8 \pi G_{N}} \, \mathbf{g}_{3} \,,
\end{equation}
where $G_{N}$ is Newton's constant and $\mathbf{g}_3$ is from \eqref{ADSASMPF}. 
Other thermodynamic variables of the field theory, the temperature ($T$) and the entropy density ($s$), can be read with the metric \eqref{METRIC} at the horizon $z_h$:
\begin{equation}\label{entropyF}
\begin{split}
T =  \left.  \frac{f'(z)}{4\pi} \sqrt{\frac{g(z)}{f(z)}} \right|_{z_{h}} \,, \quad s = \left. \frac{L^2}{4 G_{N}} \frac{h(z)}{z^2} \right|_{z_{h}} \,.
\end{split}
\end{equation}

\begin{figure}[]
\centering
     \includegraphics[width=10.0cm]{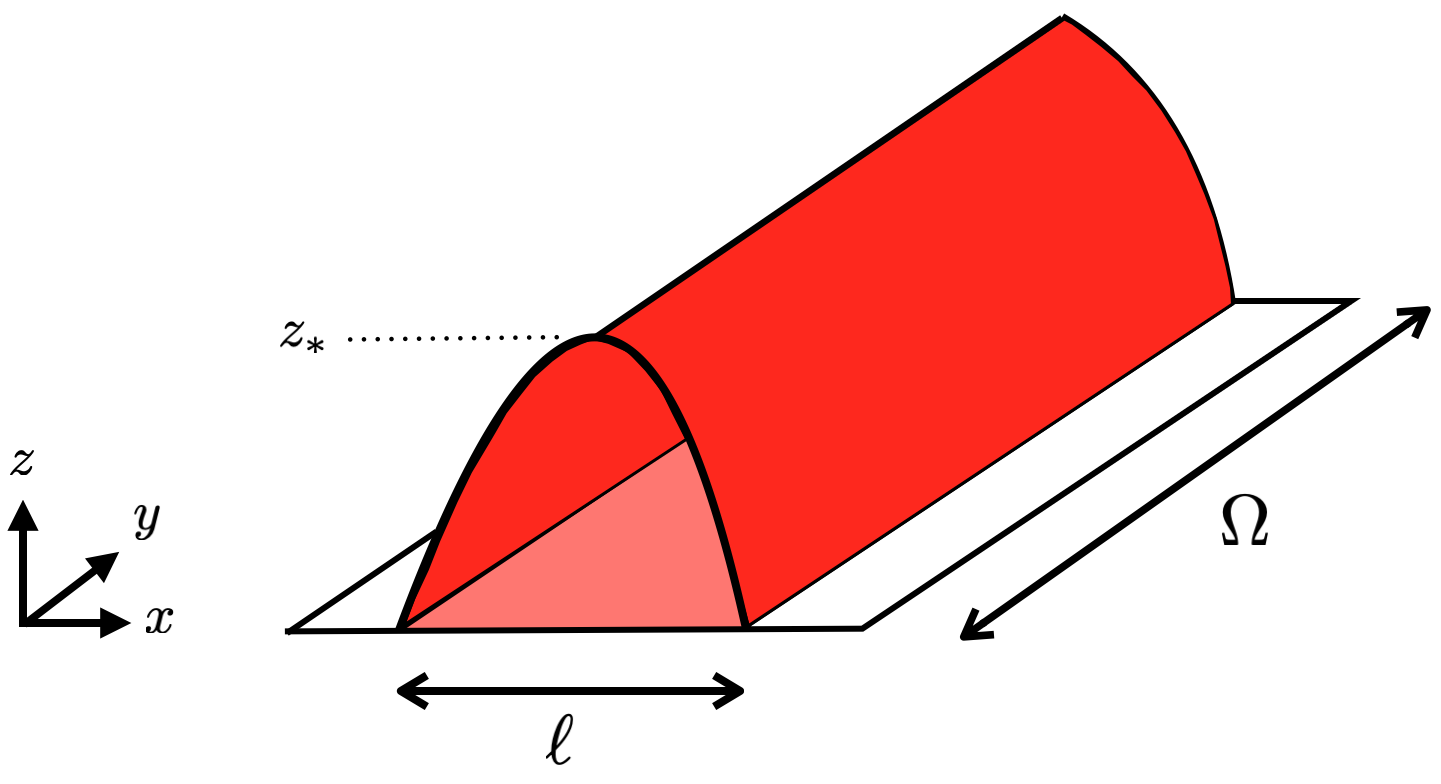}
\caption{A strip entangling region with its minimal surface (red). The strip has width $\ell$ in the $x$-direction and  length $\Omega$ in the $y$-direction. $z_{*}$ is the largest $z$ value of the minimal surface in the bulk.}\label{STRIPfig}
\end{figure}

\paragraph{Holographic entanglement entropy $S$:} 
One can study the entanglement entropy ($S$) holographically via~\cite{Ryu:2006bv,Ryu:2006ef,Lewkowycz:2013nqa}
\begin{align}\label{HEEFOR}
S = \frac{\mathcal{A}_{\text{min}}}{4G_{N}} \,,
\end{align}
where $\mathcal{A}_{\text{min}}$ is the area of the minimal surface in the bulk at a fixed $t$, which is anchored at the AdS boundary. 
In this paper, we consider the strip entangling surface. See Fig. \ref{STRIPfig}: the minimal surface $\mathcal{A}_{\text{min}}$ is expressed as a red surface with the width of the strip $\ell$ in which $z_{*}$ is the largest $z$ value of the minimal surface in the bulk.

With the metric \eqref{METRIC}, geometric quantities for \eqref{HEEFOR}, $\ell$ and $\mathcal{A}_{\text{min}}$, can be obtained as 
\begin{align}\label{HEEFORSTIP}
\begin{split}
\ell  &=\, 2 \int^{z_{*}}_{0} \dd z \, \frac{z^2}{z_{*}^2 \sqrt{\frac{h(z)^2}{h(z_{*})^2} - \frac{z^4}{z_{*}^4}}} \frac{1}{\sqrt{g(z)\,h(z)}}  \,, \\
\mathcal{A}_{\text{min}} &=\,  2L^2 \, \Omega \, \int^{z_{*}}_{\epsilon} \dd z {\frac{1}{z^2 \sqrt{1- \frac{z^4}{z_{*}^4}\frac{h(z_{*})^{2}}{h(z)^{2}}  }}} \sqrt{\frac{h(z)}{g(z)}} \,,
\end{split}
\end{align}
where the lower endpoint in $\mathcal{A}_{\text{min}}$ is a cutoff $z=\epsilon$ dual to the UV divergence in holography.

In order to isolate the UV divergence, it may be useful to split the integrand of $\mathcal{A}_{\text{min}}$ in \eqref{HEEFORSTIP} into two parts as
\begin{align}\label{USINT}
\begin{split}
\int^{z_{*}}_{\epsilon} \dd z {\frac{1}{z^2 \sqrt{1- \frac{z^4}{z_{*}^4} \frac{h(z_{*})^{2}}{h(z)^{2}} }}} \sqrt{\frac{h(z)}{g(z)}} \,=\, \int^{z_{*}}_{\epsilon} \frac{\dd z}{z^2} \,+\, \int^{z_{*}}_{\epsilon\rightarrow0} \frac{\dd z}{z^2} \left( {\frac{1}{\sqrt{1- \frac{z^4}{z_{*}^4} \frac{h(z_{*})^{2}}{h(z)^{2}}}}} \sqrt{\frac{h(z)}{g(z)}} - 1 \right)\,,
\end{split}
\end{align}
where the UV divergence can be easily evaluated from the first term and we can let $\epsilon\rightarrow0$ in the second integral since the second term is finite.

Using \eqref{USINT} with the change of variable $u=z/z_{*}$, one can express the minimal surface in  \eqref{HEEFORSTIP} as 
\begin{align}\label{HEEFORSTIP2}
\begin{split}
\mathcal{A}_{\text{min}} &=\,  2L^2 \, \Omega \,  \left[ \frac{1}{\epsilon} \,-\, \frac{1}{z_{*}} \,+\, \frac{h(z_{*})}{2 \, z_{*}^2} \,\ell \,+\, \frac{1}{z_{*}} \int^{1}_{0} \frac{\dd u}{u^2} \left( \sqrt{1-u^4\,\frac{h(z_{*})^2}{h(z_{*}\,u)^2}}\sqrt{\frac{h(z_{*}\, u)}{g(z_{*}\, u)}} \,-\, 1 \right) \right] \,,
\end{split}
\end{align}
with the strip width
\begin{align}\label{HEEFORSTIP33}
\begin{split}
\ell  &=\, 2 z_{*} \int^{1}_{0} \dd u \, \frac{u^2}{\sqrt{\frac{h(z_{*}\,u)^2}{h(z_{*})^2} - u^4}} \frac{1}{\sqrt{g(z_{*}\,u) \, h(z_{*}\,u)}}  \,.
\end{split}
\end{align}
Then, collecting $1/z_{*}$ terms in \eqref{HEEFORSTIP2}, one can find the entanglement entropy \eqref{HEEFOR} as
\begin{equation}\label{HEEFORFIN}
\begin{split}
S  = \frac{L^2 \Omega}{2G_{N}} \left[ \frac{1}{\epsilon} +\frac{h(z_{*})}{2 \, z_{*}^2}\,\ell + \frac{C(z_{*})}{z_{*}} \right] \,,
\end{split}
\end{equation}
with the dimensionless coefficient
\begin{equation}\label{HEEFORFIN2}
\begin{split}
C(z_{*}) := -1 + \int^{1}_{0} \frac{\dd u}{u^2} \left( \sqrt{1-u^4\,\frac{h(z_{*})^2}{h(z_{*}\,u)^2}}\sqrt{\frac{h(z_{*}\, u)}{g(z_{*}\, u)}} \,-\, 1 \right) \,.
\end{split}
\end{equation}
Note that both \eqref{HEEFORSTIP33} and \eqref{HEEFORFIN2} can be solved analytically for the case of $g(z_{*} \,u)=h(z_{*}\,u)=1$, which produces the entanglement entropy of the pure AdS geometry,  $S_{CFT}$, as
\begin{equation}\label{HEEFORFIN3}
\begin{split}
   S_{\text{CFT}} =  \frac{L^2 \Omega}{2G_{N}} \left[ \frac{1}{\epsilon}  - \frac{2\pi}{\ell} \left(\frac{\Gamma(\frac{3}{4})}{\Gamma(\frac{1}{4})}\right)^{2}    \right] \,.
\end{split}
\end{equation}

\paragraph{Holographic entanglement density $\sigma$:} 
Using \eqref{HEEFORFIN}-\eqref{HEEFORFIN3}, the entanglement  density ($\sigma$) \eqref{SIGFORF} is defined as
\begin{equation}\label{EEDFOR}
\begin{split}
\sigma  :=  \frac{S - S_{\text{CFT}}}{V}  \,=\,  \frac{L^2}{4G_{N}} \left[ \frac{h(z_{*})}{z_{*}^2}  + \frac{C(z_{*})}{z_{*}}\frac{2}{\ell}  + \frac{4\pi}{\ell^2} \left(\frac{\Gamma(\frac{3}{4})}{\Gamma(\frac{1}{4})}\right)^{2}    \right]  \,,
\end{split}
\end{equation}
where the UV divergence is canceled out. $V$ is the volume of the entangling region\footnote{For AdS$_{d>4}$, $V$ is the volume for higher dimensions~\cite{Gushterov:2017vnr,Erdmenger:2017pfh,Giataganas:2021jbj}.}, defined as $V := \ell \, \Omega$, thus $\sigma$ in \eqref{EEDFOR} may specify the ``density" of the entanglement entropy. 

The entanglement density \eqref{EEDFOR} is valid for any $\ell(z_{*})$ via \eqref{HEEFORSTIP33}.
Since the minimal surface cannot penetrate the horizon, one can notice that $\ell(z_{*})$ is evaluated in the range $0 < z_{*} <z_{h}$.
As explained in the introduction, the entanglement entropy can exhibit interesting features (FLEE and the area theorem) \eqref{SIGFORFfin} in the small or large entangling region that can be obtained as
\begin{equation}\label{SLLITs}
\begin{cases}
 z_{*}/z_{h}\,\rightarrow\,0 : &\,\,\, \text{Small entangling region}\,\,\,\,\,(\tilde{\ell}\ll1)  \,, \\
 z_{*}/z_{h}\,\rightarrow\,1 : &\,\,\, \text{Large entangling region}\,\,\,\,\,(\tilde{\ell}\gg1) \,,
\end{cases}
\end{equation}
where $\tilde{\ell}:=\ell/z_{h}$ defined in \eqref{elllfor}.

\paragraph{The first law of entanglement entropy (FLEE):}
For $\tilde{\ell}\ll1$, one may expand the integrand in \eqref{HEEFORFIN2} in the small $z_{*}/z_{h}$ expansion and integrate it order-by-order, so that \eqref{EEDFOR} could also be expanded in powers of $\tilde{\ell}$ via \eqref{HEEFORSTIP33}.\footnote{We also expand the integrand in $\ell$ of \eqref{HEEFORSTIP33} in small $ z_{*}/z_{h}$ and change the variable from $z_{*}/z_{h}$ to $\tilde{\ell}$.}
Then, if the geometries allow FLEE,  $\sigma$ will be expressed as 
\begin{equation}\label{FLEESIGMA}
\begin{split}
\sigma = \langle T_{tt} \rangle \, T_{\text{ent}}^{-1}  \,+\, \dots \,,
\end{split}
\end{equation}
where $\langle T_{tt} \rangle$ is \eqref{EDFTF}, and $T_{\text{ent}}$  \eqref{TENT}. In the following section, we will explicitly perform the small $z_{*}/z_{h}$ analysis when the black holes allow an analytic background geometry.

\paragraph{The area theorem:}
For $\tilde{\ell}\gg1$, $z_{*} \rightarrow z_{h}$, $\sigma$ in \eqref{EEDFOR} becomes
\begin{equation}\label{ATVSIGMA}
\begin{split}
\sigma  \,=\,  s \,+\, \frac{s \, z_{h} \, C(z_{h})}{h(z_{h})} \frac{2}{\ell}  \,+\, \dots \,=\,  s \,-\, \Delta\alpha \, \frac{A}{V} \,+\, \dots \,,
\end{split}
\end{equation}
where we used $s$ \eqref{entropyF} and $A/V=2/\ell$. We also identify
\begin{equation}\label{}
\begin{split}
\Delta \alpha = -\frac{s \, z_{h} \, C(z_{h})}{h(z_{h})} \,,
\end{split}
\end{equation}
which can be used for the area theorem violation: $C(z_{h})>0$ (i.e., $\Delta \alpha<0$).\footnote{Note that $h(z_{h})\geq0$ if $s\geq0$ \eqref{entropyF}.}

\subsection{Dimensionless entanglement density: $\sigma/s$}
Note that $\sigma$ in \eqref{EEDFOR} has a $\mathcal{O}(1/\ell^2)$ dimension. In \cite{Gushterov:2017vnr,Erdmenger:2017pfh}, in order to render $\sigma$ dimensionless, $s$ $\eqref{entropyF}$ has been used as a scaling parameter giving
\begin{equation}\label{SCEED}
\begin{split}
\frac{\sigma}{s} \,=\,  \frac{1}{\tilde{z}_{*}^2}\,\frac{h(\tilde{z}_{*})}{h(1)}  + \frac{C(\tilde{z}_{*})}{\tilde{z}_{*}\,h(1)}\frac{2}{\tilde{\ell}}  + \frac{4\pi}{\tilde{\ell}^2\,h(1)} \left(\frac{\Gamma(\frac{3}{4})}{\Gamma(\frac{1}{4})}\right)^{2}     \,,
\end{split}
\end{equation}
where tilde variables are defined as 
\begin{align}\label{elllfor}
\begin{split}
\tilde{z}_{*} := \frac{z_{*}}{z_{h}} \,, \qquad \tilde{\ell}:= \frac{\ell}{z_{h}} = 2 \tilde{z}_{*}  \int^{1}_{0} \dd u \, \frac{u^2}{\sqrt{\frac{h(z_{*}\,u)^2}{h(z_{*})^2} - u^4}} \frac{1}{\sqrt{g(z_{*}\,u) \, h(z_{*}\,u)}} \,,
\end{split}
\end{align}
with $\ell$ \eqref{HEEFORSTIP33}. Note that the argument for $h(z_h)$ and $C(z_{*})$ have also been scaled with $z_{h}$ as $h(1)$ and $C(\tilde{z}_{*})$, respectively.
Therefore, for the given metric \eqref{METRIC}, one can study $\sigma/s$ via \eqref{SCEED} once both $C(\tilde{z}_{*})$ \eqref{HEEFORFIN2} and $\tilde{\ell}$ \eqref{elllfor} are evaluated.

In terms of $\sigma/s$, FLEE \eqref{FLEESIGMA} can be expressed as
\begin{align}\label{FLEENOUD}
\begin{split}
\text{FLEE:} \qquad \frac{\sigma}{s} =  \left( \frac{\Gamma\left( \frac{1}{4} \right)}{\Gamma\left(\frac{3}{4}\right)} \right)^2 \, \frac{\tilde{\mathbf{g}}_{3}}{32\,h(1)} \, \tilde{\ell}  \,+\, \dots  \,,
\end{split}
\end{align}
where we used \eqref{TENT} for  $T_{\text{ent}}$, \eqref{EDFTF} for $\langle T_{tt} \rangle$, and  \eqref{entropyF} for $s$. We also define $\tilde{\mathbf{g}}_{3} := \mathbf{g}_{3} \, z_{h}^3$.
One may also choose a different scaling parameter to make $\sigma$ dimensionless, however, $s$ is useful for our purpose: to check the area theorem, i.e., \eqref{SCEED} at $\tilde{\ell}\gg1$ (or $\tilde{z}_{*}\rightarrow1$)
\begin{equation}\label{ARNOUD}
\begin{split}
\text{Area theorem:} \qquad \frac{\sigma}{s}  &\,=\,  1 \,+\, \frac{C(1)}{h(1)} \frac{2}{\tilde{\ell}}  \,+\, \dots \,,
\end{split}
\end{equation}
where the area theorem is violated when $C(1)>0$.

In what follows, using holographic superconductor models and holographic axion models, we will study if FLEE \eqref{FLEENOUD} and/or the area theorem \eqref{ARNOUD} is violated or not when the $U(1)$ symmetry is broken in section \ref{sec3label}, and when the translational symmetry is broken in section \ref{sec4label}.

%
\section{Broken $U(1)$ symmetry}\label{sec3label}

\subsection{The model}
We study a holographic superconductor model based on Einstein-Maxwell theory~\cite{Hartnoll:2008vx,Hartnoll:2008kx}\footnote{For the recent development of holographic superconductors, we refer the reader to \cite{Gouteraux:2019kuy,Gouteraux:2020asq,Arean:2021brz,Ammon:2021slb,Jeong:2021wiu,Baggioli:2022aft} and references therein.}:
\begin{align} \label{action1}
\begin{split}
S & =  S_1 + S_2  =  \int \dd^4x\sqrt{-g}\left( \mathcal{L}_1 + \mathcal{L}_2  \right) \,, \\ 
\mathcal{L}_1 &= R  + 6 -\frac{1}{4} \,F^2  \,, \quad  \mathcal{L}_2 =  -|D\Phi|^2  - M^2 |\Phi|^2  \,,
\end{split}
\end{align}
where we set units such that the gravitational constant $16\pi G=1$ and the AdS radius $L=1$ for simplicity.
The action \eqref{action1} consists of two actions. $S_1$ is the Einstein-Maxwell theory composed of two fields: the metric $g_{\mu\nu}$, a $U(1)$ gauge field $A_{\mu}$ with field strength $F=\dd A$. The second action $S_2$ is for the superconducting phase, constructed by a complex scalar field $\Phi$ with the covariant derivative $D_\mu \Phi=  \left(  \nabla_\mu -i q A_\mu   \right)  \Phi$. For numerics, we set $q=3$ in this paper.

In order to study the action \eqref{action1}, we take the following ansatz for numerical convenience 
\begin{equation}\label{NormalAnsatz}
\begin{split}
&\dd  s^2 =  \frac{1}{z^2} \left[-\left(1-\frac{z}{z_{h}}\right)U(z)e^{-S(z)} \dd t^2+\frac{\dd z^2}{\left(1-\frac{z}{z_{h}}\right)U(z)}+\dd x^2 + \dd y^2 \right] \,, \\
&A=\left(1-\frac{z}{z_{h}}\right)a(z) \dd t \,, \quad  \Phi=z^{\Delta_{-}}\, \eta(z) \,,
\end{split}
\end{equation}
where $z_h$ is the horizon and the AdS boundary is at $z=0$.
Comparing \eqref{METRIC} with \eqref{NormalAnsatz}, one can find that
\begin{equation}\label{ALNORSL}
\begin{split}
f(z)=\left(1-\frac{z}{z_h}\right)U(z)e^{-S(z)} \,,\quad g(z)=\left(1-\frac{z}{z_h}\right)U(z)\,, \quad h(z)=1 \,,
\end{split}
\end{equation}
and the energy density ($\langle T_{tt} \rangle$), the temperature ($T$), and the entropy density ($s$) for the action \eqref{action1} can be computed via \eqref{ADSASMPF}-\eqref{entropyF}.

In order for the bulk geometry asymptotic to the AdS spacetime near the boundary ($z=0$), we impose the boundary condition for the metric as $U(0)=1$, $S(0)=0$.
It also turns out that the boundary behavior of the matter fields $A_t$, $\Phi$ is
\begin{align} \label{bdexpan33}
\begin{split}
A_t = \mu \,-\, \rho  \,z \,+\, \dots \,, \qquad
\Phi = \Phi^{(-)} \, z^{\Delta_{-}} \,+\, \Phi^{(+)} \, z^{\Delta_{+}} \,+\, \dots  \,,
\end{split}
\end{align}
where $\Delta_{\pm} = \frac{3}{2} \pm \sqrt{\frac{9}{4} + M^2}$. According to the holographic dictionary, $\mu$ is interpreted as the chemical potential and $\rho$ is the charge density. In the asymptotic form of $\Phi$, $\Phi^{(-)}$ is the source and $\Phi^{+}$ is the condensate. Then, as the boundary condition for the superconducting phase, we set the source to be zero, $\Phi^{(-)}=0$, to describe the spontaneous symmetry breaking. Thus, one can have a superconducting phase with $\Phi^{(+)} \neq 0$ and a normal phase with $\Phi = 0$.
Note that, from the ansatz \eqref{NormalAnsatz}, one can easily read off the chemical potential $\mu$ via $\mu=a(0)$ and the source $\Phi^{(-)}$ via $\Phi^{(-)}=\eta(0)$.

\subsection{Normal phase: a review}
Let us first review the normal phase ($\Phi$ = 0) \cite{Gushterov:2017vnr}, $S=S_1$ in \eqref{action1}. In the normal phase, one can find the analytic solution as
\begin{align} \label{}
\begin{split}
U(z) = 1+\frac{z}{z_h}+\frac{z^2}{z_h^2} -\frac{\mu^2}{4}\frac{z^3}{z_h}, \quad  S(z) = 0,\quad  a(z) = \mu, \quad \eta(z) = 0 \,,
\end{split}
\end{align}
which corresponds to
\begin{equation}\label{LAMSOBG}
\begin{split}
f(z) = g(z) = 1 - \mathbf{g}_3 \, z^3 + \frac{\mu^2}{4 z_h^2} z^4   \,,\quad \mathbf{g}_3 \, = \frac{1}{z_h^{3}} \left( 1 + \frac{\mu^2 z_h^2}{4} \right) \,,
\end{split}
\end{equation}
via \eqref{ALNORSL}. Note that the same notation $\mathbf{g}_3$ in \eqref{ADSASMPF} is used here.
The temperature $T$ \eqref{entropyF} reads
\begin{align}\label{TEMNORMAL}
\begin{split}
T = \frac{3}{4\pi z_h} - \frac{\mu^2 z_h}{16 \pi}   \,.
\end{split}
\end{align}

\paragraph{The entanglement density for normal phase:}
In order to evaluate \eqref{SCEED}, we need to identify $g(z_{*} \, u)$ as
\begin{align}\label{gfnRN}
\begin{split}
g(z_{*} \, u) = 1 - \tilde{\mathbf{g}}_{3} \, \tilde{z}_{*}^3 \, u^3 + \frac{\tilde{\mu}^2}{4} \tilde{z}_{*}^4 \, u^4 \,, 
\end{split}
\end{align}
which is the only input function for both $C(\tilde{z}_{*})$ in \eqref{HEEFORFIN2} and $\tilde{\ell}$ in \eqref{elllfor}.
In \eqref{gfnRN}, we also used $\tilde{\mathbf{g}}_{3} := \mathbf{g}_{3} \, z_{h}^3$,\, $\tilde{\mu}:=\mu \, z_{h}$.

We want to fix the chemical potential for the normal phase, so $\sigma/s$ should be expressed in terms of $\ell$ and $T$ at fixed $\mu$: i.e., ($\ell\,\mu, \, T/\mu$).
For this purpose, we can use
\begin{equation}\label{EDERS348}
\begin{split}
 \ell \, \mu \,=\, \tilde{\ell} \, \tilde{\mu}   \,, \qquad  \frac{T}{\mu} \,=\, \frac{1}{4\pi \tilde{\mu}}\left( 3 - \frac{\tilde{\mu}^2}{4} \right),
\end{split}
\end{equation}
where $\tilde{\ell}:=\ell/z_{h}$ \eqref{elllfor} and $T$ from \eqref{TEMNORMAL}.
Note that once \eqref{gfnRN} is being used, $\sigma/s$ \eqref{SCEED} can be a function of ($\tilde{z}_{*}, \, \tilde{\mu}$) which can be expressed further in terms of  ($\tilde{\ell}, \, \tilde{\mu}$) via \eqref{elllfor}. 
Furthermore, solving the relations \eqref{EDERS348} we can find the expression of ($\tilde{\ell}, \, \tilde{\mu}$) in terms of ($\ell\,\mu, \, T/\mu$). Thus, using \eqref{EDERS348}, $\sigma/s$ \eqref{SCEED} can be evaluated at given ($\ell \, \mu, \, T/\mu$).

\begin{figure}[]
 \centering
      \subfigure[$\sigma/s$ vs $\ell \, \mu$]
     {\includegraphics[width=7.0cm]{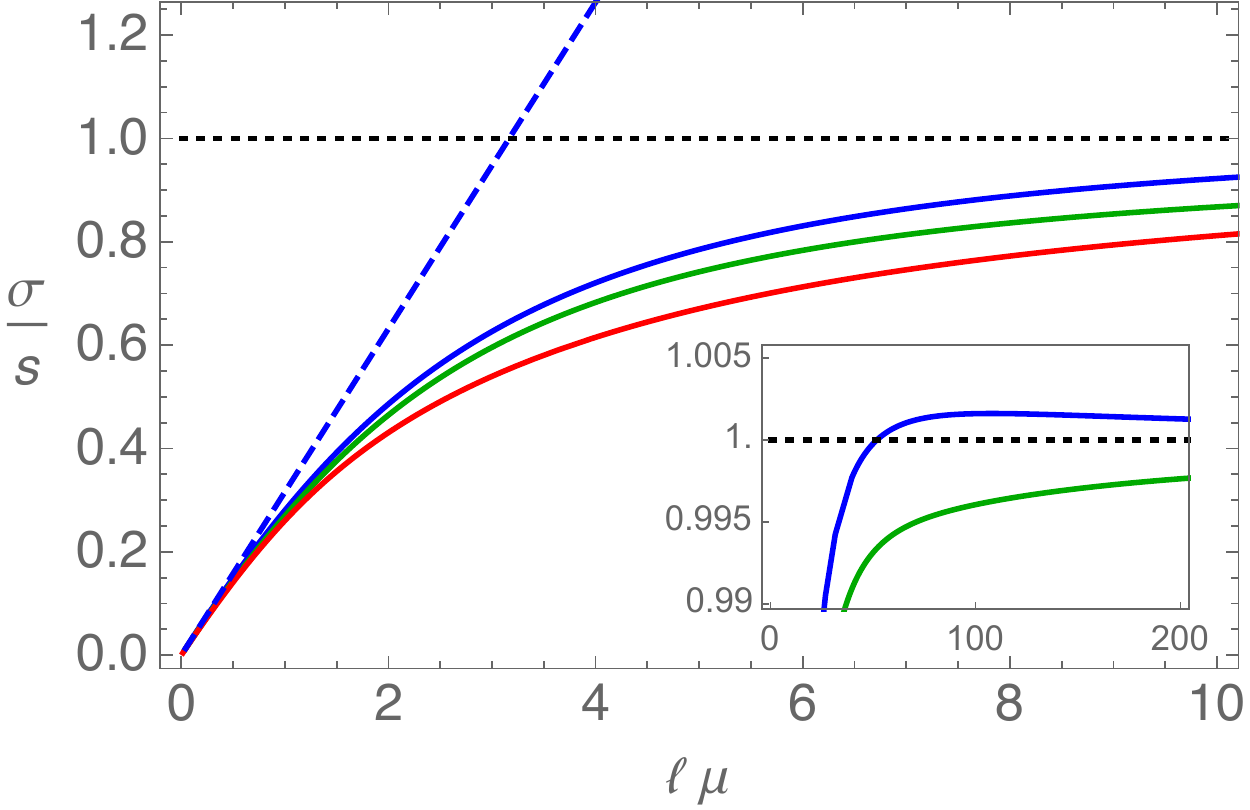}\label{FIGRN1}} 
      \subfigure[$C(1)$ vs $T/\mu$]
     {\includegraphics[width=7.7cm]{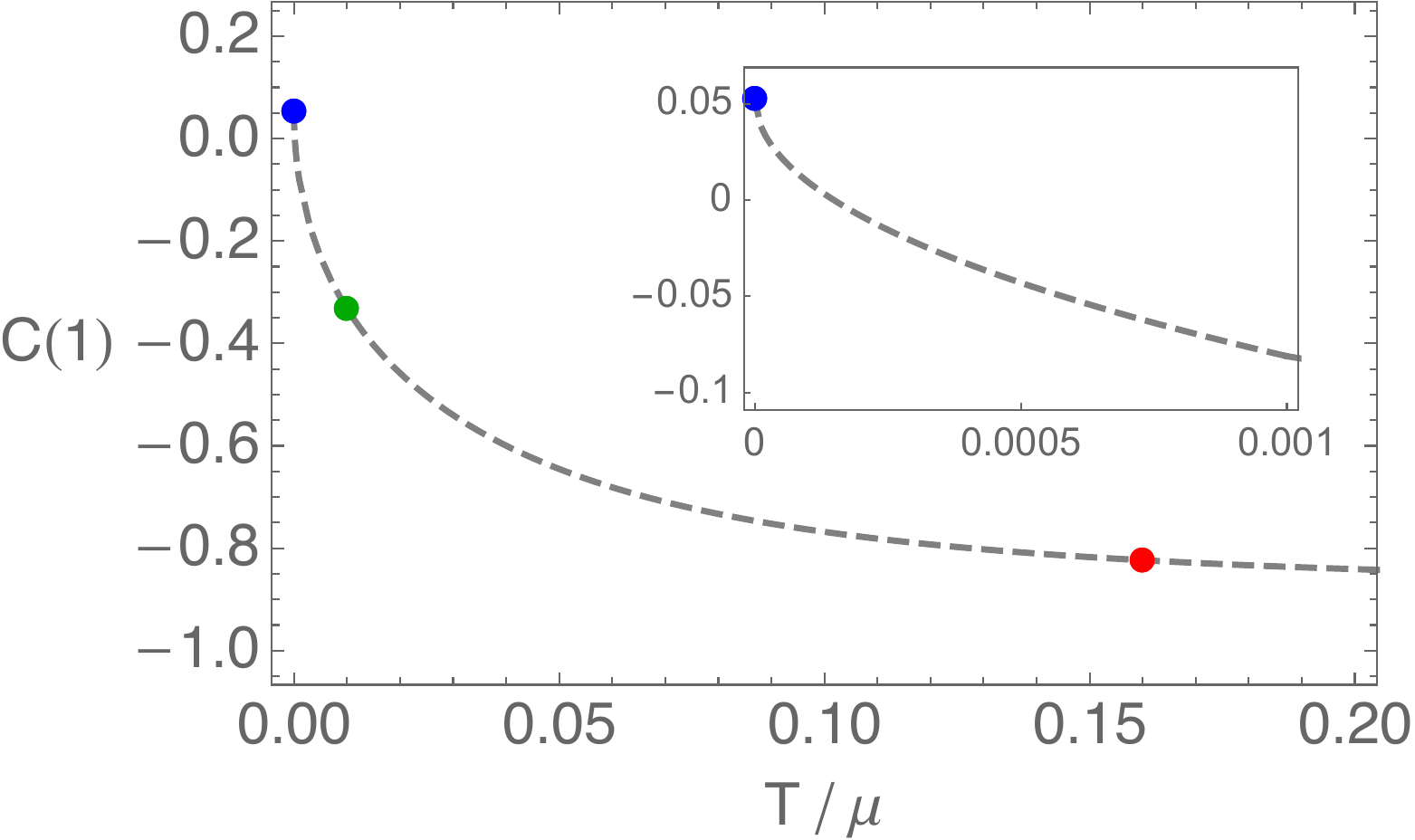} \label{FIGRN2}} 
          \caption{Entanglement density of the normal phase. $\textbf{Left:}$ $\sigma/s$ (solid lines) at $T/\mu=(0.16, \,0.01, \,0)$ (red, green, blue). Black dotted line is a guide line for $\sigma/s=1$ and the blue dashed line is \eqref{STE2DS}. The inset shows the large $\ell\,\mu$ behavior. $\textbf{Right:}$ $C(1)$ (dashed line). Dots correspond to data used in Fig. \ref{FIGRN1}. The inset shows the low $T/\mu$ behavior: $C(1)>0$ at low $T$. } \label{FIGRN}
\end{figure}

At given ($\ell \, \mu, \, T/\mu$), we investigate $\sigma/s$ for the normal phase and found that 
\begin{equation}\label{NPRET}
\begin{split}
\text{Normal phase:} \qquad \text{FLEE is obeyed,} \quad \text{Area theorem is violated at low $T$} \,,
\end{split}
\end{equation}
which can be seen in Fig. \ref{FIGRN}.
In Fig. \ref{FIGRN1}, we display $\sigma/s$ with different $T/\mu=(0.16, \,0.01, \,0)$ (red, green, blue).
One can find that FLEE \eqref{FLEENOUD} is obeyed for all $T$: $\sigma \sim \ell$ at low $\ell$, e.g, see the blue dashed  line ($T/\mu=0$ case) in Fig. \ref{FIGRN1}.

For the area theorem \eqref{ARNOUD}, we need to study the large $\ell$ behavior. From the inset in Fig. \ref{FIGRN1}, one can see that the area theorem is violated at low $T$: e.g., $\sigma/s\rightarrow 1^{+}$ for the $T/\mu=0$ case (the blue  solid line). The violation of the area theorem can also be easily checked by $C(1)>0$ at low $T$ in Fig. \ref{FIGRN2}.

\paragraph{FLEE of the charged black holes:}
We close this subsection with the analysis on how to obtain \eqref{FLEENOUD} for the charge black hole $S=S_1$ in \eqref{action1}. Our analytic result is complementary to the numerical results in \cite{Gushterov:2017vnr}.\footnote{One can find the neutral case in \cite{Erdmenger:2017pfh}.}
 
In order to study FLEE, we need to consider the $\tilde{\ell}\ll1$ limit or equivalently $\tilde{z}_{*}\rightarrow0$ limit in \eqref{gfnRN}. 
First, for $\tilde{\ell}$ \eqref{elllfor}, one can find its leading behavior by considering $g(z_{*} \, u)=1$ as 
\begin{equation}\label{smallLex}
\begin{split}
\tilde{\ell}  \, = \,  2\sqrt{\pi} \, \frac{\Gamma\left( \frac{3}{4} \right)}{\Gamma\left( \frac{1}{4} \right)}  \, \tilde{z}_{*}  \, + \, \dots \,.
\end{split}
\end{equation}
Moreover, it is useful to express $C$ \eqref{HEEFORFIN2} up to its sub-leading order by considering $g(z_{*} \, u)=1 - \tilde{\mathbf{g}}_{3} \, \tilde{z}_{*}^3 \, u^3$ in \eqref{gfnRN}, which is
\begin{equation}\label{smallC}
\begin{split}
C(\tilde{z}_{*}) &= -\frac{2\sqrt{\pi}}{3}\frac{\Gamma\left( \frac{7}{4} \right)}{\Gamma\left( \frac{5}{4} \right)} \,+\, \frac{\pi}{16} \, \tilde{\mathbf{g}}_{3} \, \tilde{z}_{*}^3 \,+\, \dots \,.
\end{split}
\end{equation}

Then, using \eqref{smallLex}, $\sigma/s$ \eqref{SCEED} can be expressed as  
\begin{equation}\label{STE1DS}
\begin{split}
\frac{\sigma}{s} \,=\, \frac{1}{\tilde{z}_{*}^2}\left[2  + \frac{3}{\sqrt{\pi}}\frac{\Gamma\left(\frac{5}{4}\right)}{\Gamma\left(\frac{7}{4}\right)} \, C(\tilde{z}_{*}) \, \right]  \,+\, \dots \,, 
\end{split}
\end{equation}
where the constant $2$ in \eqref{STE1DS} comes from the combination between the first and third terms in \eqref{SCEED}.\footnote{Recall that $h(z)=1$ in \eqref{ALNORSL}.}
Moreover, plugging $C$ \eqref{smallC} into \eqref{STE1DS}, the constant term in \eqref{STE1DS}, $2$, will be eliminated by the leading term of $C$ in \eqref{smallC}, $\frac{-2\sqrt{\pi}}{3}\frac{\Gamma\left( \frac{7}{4} \right)}{\Gamma\left( \frac{5}{4} \right)}$.
Then $\sigma/s$ can show FLEE \eqref{FLEENOUD} by the sub-leading term of $C$ in \eqref{smallC} as
\begin{equation}\label{STE2DS}
\begin{split}
\frac{\sigma}{s} \,=\, \frac{3\sqrt{\pi}}{16} \, \frac{\Gamma\left(\frac{5}{4}\right)}{\Gamma\left(\frac{7}{4}\right)} \, \tilde{\mathbf{g}}_{3} \, \tilde{z}_{*}   \,=\, \left( \frac{\Gamma\left( \frac{1}{4} \right)}{\Gamma\left(\frac{3}{4}\right)} \right)^2 \, \frac{\tilde{\mathbf{g}}_{3}}{32} \, \tilde{\ell}  \,=\, \left( \frac{\Gamma\left( \frac{1}{4} \right)}{\Gamma\left(\frac{3}{4}\right)} \right)^2 \, \frac{\tilde{\mathbf{g}}_{3}}{32 \, \tilde{\mu}} \, (\ell\,\mu)  \,,   
\end{split}
\end{equation}
where we used \eqref{smallLex} in the second equality and $\ell \, \mu \,=\, \tilde{\ell} \, \tilde{\mu}$ \eqref{EDERS348} in the last equality.

\subsection{Superconducting phase}
Next, let us study $\sigma/s$ \eqref{SCEED} of the superconducting phase ($\Phi\neq0$), $S=S_1+S_2$ in \eqref{action1}. 
Solving equations of motion from \eqref{action1}, one can find the numerical solutions $g(z_{*} \, u)$ for $T<T_c$ where $T_c$ could be identified by the temperature at which the condensate $\Phi^{(+)}$ starts to be finite. Then, with numerical solutions, one can evaluate $C(\tilde{z}_{*})$ in \eqref{HEEFORFIN2}, $\tilde{\ell}$ in \eqref{elllfor} so that $\sigma/s$ in \eqref{SCEED} for the superconducting phase.

\begin{figure}[]
 \centering
      \subfigure[$\sigma/s$ vs $\ell \, \mu$]
     {\includegraphics[width=7.1cm]{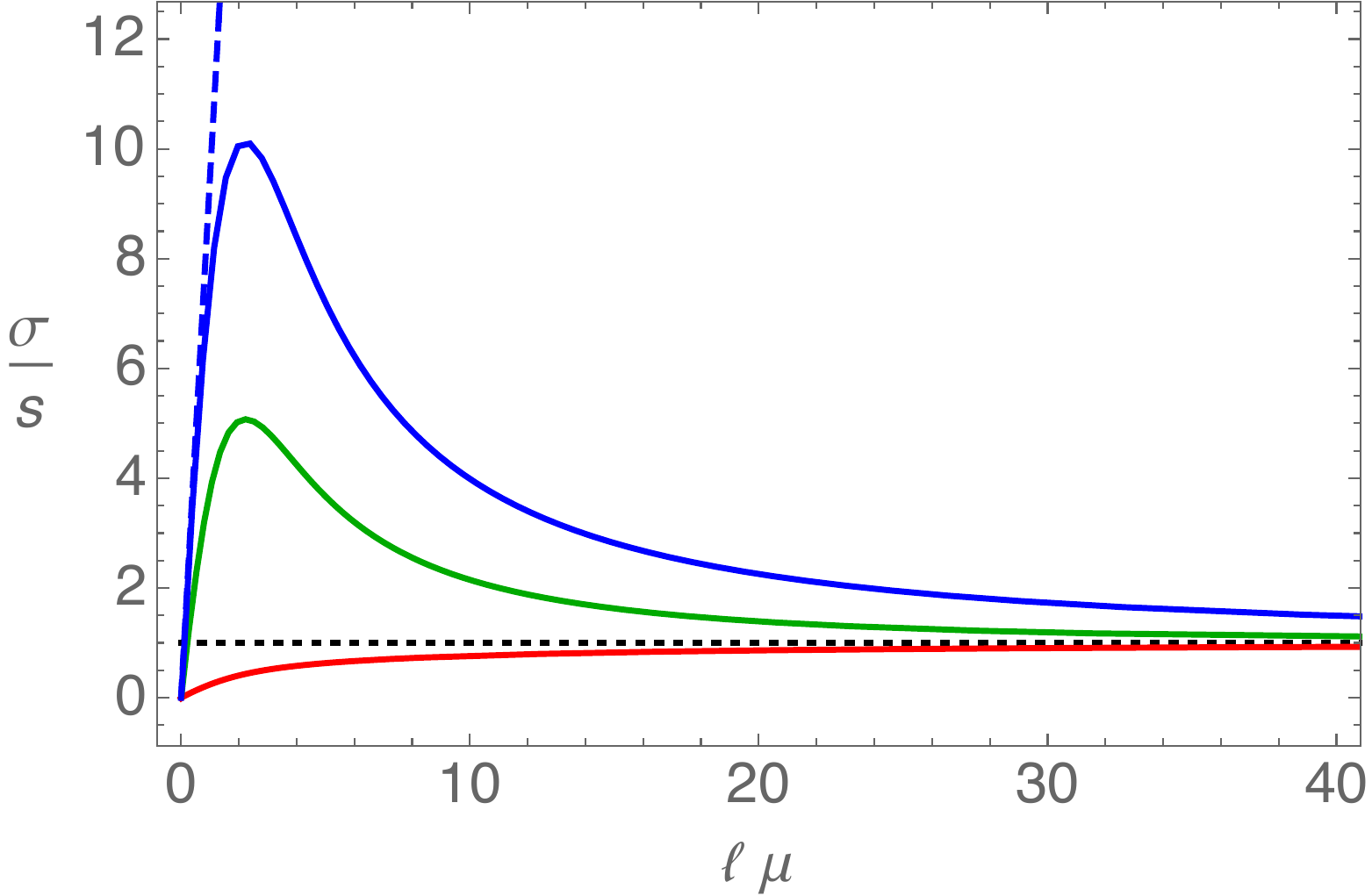}\label{sc0FIGSC1}} 
      \subfigure[$C(1)$ vs $T / T_c$]
     {\includegraphics[width=7.6cm]{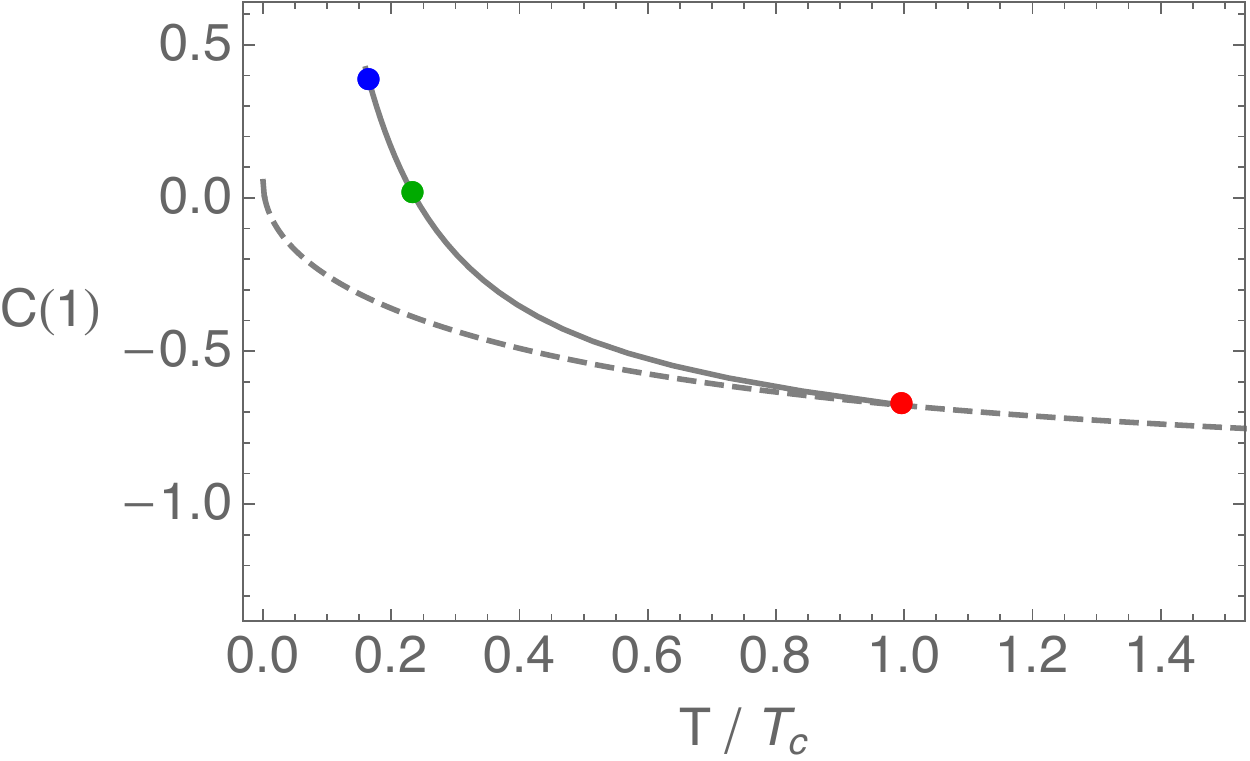} \label{sc0FIGSC2}} 
          \caption{Entanglement density of the superconducting phase at $M^2=0$. (For $M^2 >0$ we obtained qualitatively the same plots.) $\textbf{Left:}$ $\sigma/s$ (solid lines) at $T/T_c=(1, \,0.23, \,0.16)$ (red, green, blue). The black dotted line is the guide line for $\sigma/s=1$ and the blue dashed line is \eqref{FLEENOUD}.  $\textbf{Right:}$ $C(1)$ for the superconducting phase (solid line), the normal phase (dashed line). Dots correspond to data used in Fig. \ref{sc0FIGSC1}.  }
 \label{sc0FIGSC}
\end{figure}

\begin{figure}[]
 \centering
      \subfigure[$\sigma/s$ vs $\ell \, \mu$]
     {\includegraphics[width=7.3cm]{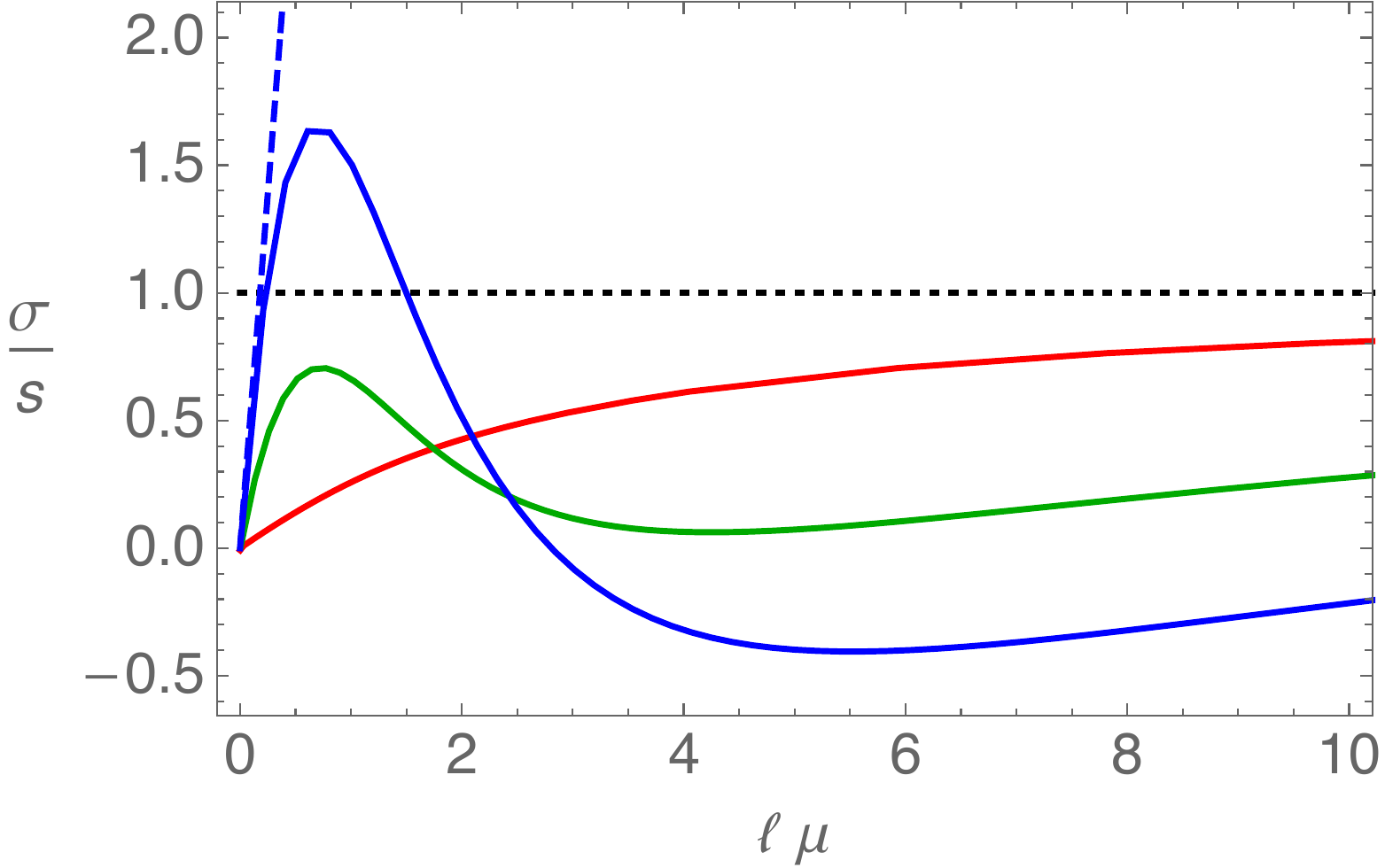}\label{scFIGSC1}} 
      \subfigure[$C(1)$ vs $T / T_c$]
     {\includegraphics[width=7.5cm]{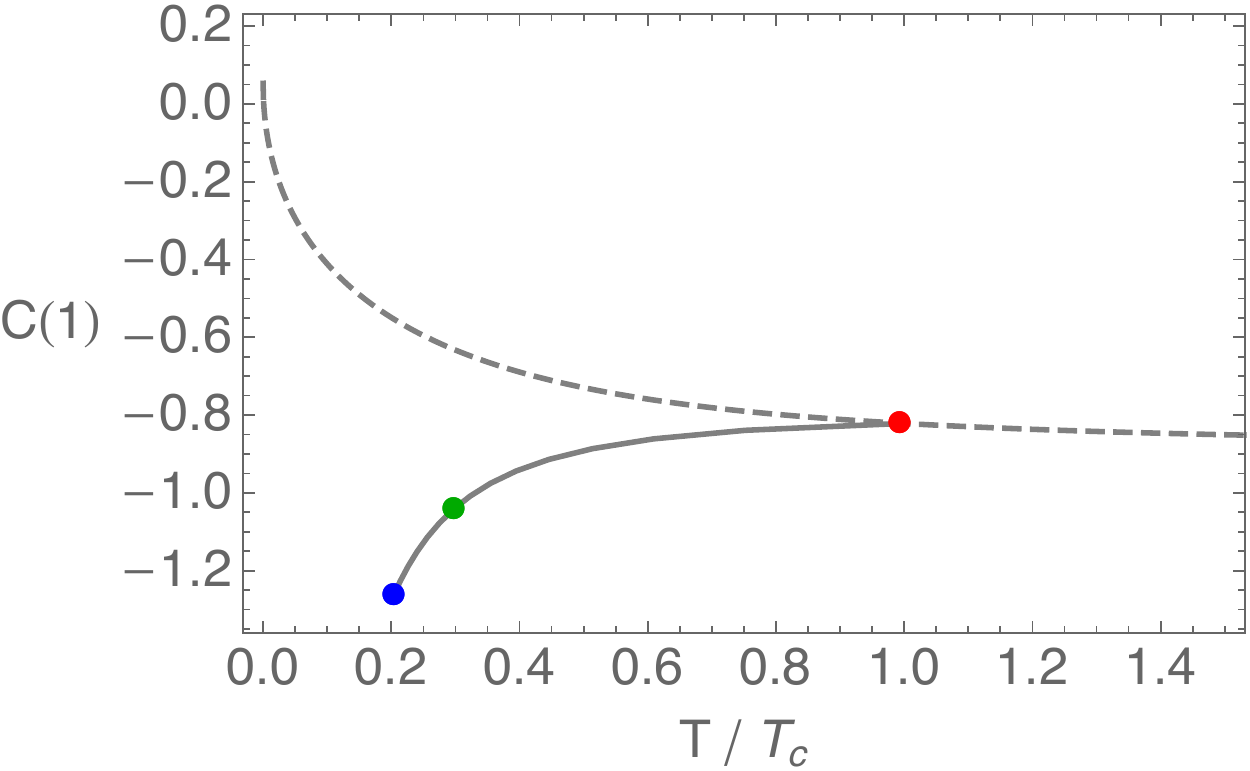} \label{scFIGSC2}} 
          \caption{Entanglement density of the superconducting phase at $M^2=-2$.  $\textbf{Left:}$ $\sigma/s$ (solid lines) at $T/T_c=(1, \,0.3, \,0.2)$ (red, green, blue). The black dotted line is the guide line for $\sigma/s=1$ and the blue dashed line is \eqref{FLEENOUD}.  $\textbf{Right:}$ $C(1)$ for the superconducting phase (solid line), the normal phase (dashed line). Dots correspond to data used in Fig. \ref{scFIGSC1}.  } \label{scFIGSC}
\end{figure}

In particular, we make the plot of $\sigma/s$ for two cases: i)  $M^2=0$ in Fig. \ref{sc0FIGSC}; ii) $M^2=-2$ in Fig. \ref{scFIGSC} in order to examine the mass ($M^2$) dependence of $\sigma/s$.\footnote{See also \cite{Yang:2019gce} for the $M^2$ dependence of another quantum information quantity: the holographic complexity.} These are the representative examples for $M^2\geq0$ and $M^2<0$, respectively.\footnote{We also checked the $M^2=2$ case. However, the result of $M^2=2$ is qualitatively the same as $M^2=0$.}

\paragraph{The entanglement density for the superconducting phase:} 
For $M^2\geq0$, we found that $\sigma/s$ for superconductors is qualitatively similar to the normal phase \eqref{NPRET}, i.e., FLEE is obeyed (see Fig. \ref{sc0FIGSC1}), while the area theorem is violated at low $T$ (see Fig. \ref{sc0FIGSC2}).\footnote{Note that \eqref{FLEENOUD} for the superconducting phase can be evaluated with  $\tilde{\ell}\,=\,\ell \, \mu / \tilde{\mu}$ \eqref{EDERS348} with $\tilde{\mathbf{g}}_{3}$ from the numerical solution of $g(z)$ \eqref{ADSASMPF} and $\tilde{\mu}$ from $a(0)$ \eqref{NormalAnsatz}. Note also that we set $z_{h}=1$ for numerics, thus $\mu=\tilde{\mu}$ and $\mathbf{g}_{3}=\tilde{\mathbf{g}}_{3}$.}

On the other hand, the $M^2<0$ case produces the different result from the $M^2\geq0$ case: both FLEE and the area theorem are obeyed. 
For FLEE, see Fig. \ref{scFIGSC1} in which $\sigma \sim \ell$ at low $\ell$ for all $T<T_c$.
For the area theorem, see Fig. \ref{scFIGSC2}: $C(1)$ for superconducting phase (solid line) may always be negative, indicating the area theorem is obeyed for all $T<T_c$.


Note that $C(1)$ for the superconducting phase in Fig. \ref{scFIGSC2} does not reach to $T/T_c=0$ because of the instability in our numerics. However, from the large $\ell\,\mu$ behavior of $\sigma/s$ in Fig. \ref{scFIGSC1}, it may be expected that the area theorem for the superconducting phase is obeyed even at lower $T/T_c$ because, in the large $\ell\,\mu$ regime, the value of $\sigma/s$ is smaller at lower $T$, e.g., at $\ell \, \mu=10$, $\sigma/s$ is decreasing from $0.8$ ($T/T_c=1$, red) to $-0.2$ ($T/T_c=0.2$, blue).
This implies that $C(1)$ in \eqref{ARNOUD} is decreasing as $T$ is lowered, which is consistent with Fig. \ref{scFIGSC2}. Thus, $\sigma/s$ may be approaching $1^{-}$ for all $T<T_c$.

In summary, for holographic superconductors, the obedience/violation of FLEE/area theorem depends on the sign of $M^2$:
\begin{align}\label{SCPRETM2M2}
\begin{split}
{\text{Superconducting phase with   }} & M^2\geq0: \,\, \small{\text{FLEE is obeyed,} \,\,\, \text{Area theorem is violated at small $T$}}\,, \\
{\text{Superconducting phase with   }} & M^2<0: \,\, \small{\text{FLEE is obeyed,} \,\,\, \text{Area theorem is obeyed}}\,.
\end{split}
\end{align}
If the area theorem were obeyed in the superconducting phase, the violation of the area theorem would have played the role of distinguishing the normal phase from  the superconducting phase. However, this is not the case. Instead, it seems that the violation of the area theorem may classify the superconducting phases: one class with  $M^2 \geq0 $ and the other with $M^2 < 0 $. Noting that $M^2$ is related to the material properties such as the critical temperature $T_c$ in condensed matter systems we may say $M^2$ can identify different superconducting materials. 

For a summary of FLEE/area theorem, see Table. \ref{SUM1}.
\begin{table}[]
\begin{tabular}{| >{\centering\arraybackslash}m{1.75in} | >{\centering\arraybackslash}m{1.83in} | >{\centering\arraybackslash}m{1.85in}  |}
\hline
   U(1) symmetry   & FLEE   & Area theorem  \\ 
 \hline
 \hline
     \small{Normal phase}   &  \small{Obeyed}      &  \small{Violated}      \\
 \hline
  \hline
   \small{Superconducting phase ($M^2\geq0$)}   &  \small{Obeyed}     &       \small{Violated}     \\
 \hline
   \small{Superconducting phase ($M^2<0$)}  &  \small{Obeyed}     &       \small{Obeyed}          \\
 \hline
\end{tabular}
\caption{FLEE/area theorem with broken $U(1)$ symmetry. The area theorem is violated at low $T$.}\label{SUM1}
\end{table}
%

%
\section{Broken translational symmetry}\label{sec4label}
\subsection{The model}
We consider the Einstein-Maxwell-Axion model~\cite{Baggioli:2021xuv} as 
\begin{equation}\label{GENACa}
\begin{split}
S = \int \dd^4x \sqrt{-g} \,\left[ R + 6 - \frac{1}{4}F^2 - X^N \right] \,,
\end{split}
\end{equation}
which is constructed with $S_{1}$ in \eqref{action1} by adding an additional scalar field $\varphi_i$ called the axion field 
\begin{align}\label{BSASC}
\begin{split}
X := \frac{1}{2}\sum_{i=1}^{2}\left(\partial\varphi_i\right)^2 \,, \qquad \varphi_i = m \, x^i,
\end{split}
\end{align}
where $m$ denotes the strength of the translational symmetry breaking.

The action \eqref{GENACa} allows analytic background solutions as
\begin{equation}\label{bgmora} 
\begin{split}
\dd s^2 =  \frac{1}{z^2}\left[ -f(z)\, \dd t^2 +  \frac{1}{g(z)} \, \dd z^2  + \dd \vec{x}^2 \right]  \,, \quad A = A_{t}(z) \, \dd t \,,
\end{split}
\end{equation}
with 
\begin{equation}\label{BGMa}
\begin{split}
f(z) = g(z) = 1 - \mathbf{g}_3 z^3 + \frac{\mu^2}{4 z_h^2} z^4 + \frac{m^{2N}}{2(2N-3)} z^{2N}  \,,\quad  A_{t}(z) = \mu\left( 1- \frac{z}{z_h} \right) \,,
\end{split}
\end{equation}
where it becomes \eqref{LAMSOBG} at $m=0$ and $\mathbf{g}_3$ is determined by $f(z_h)=0$ as
\begin{equation}\label{BGMa2}
\begin{split}
\mathbf{g}_3 = \frac{1}{z_h^{3}} \left( 1 + \frac{\mu^2 z_h^2}{4} + \frac{m^{2N} z_h^{2N}}{2(2N-3)}  \right) \,,
\end{split}
\end{equation}
and the temperature $T$ \eqref{entropyF} is
\begin{align}\label{THMERRESULTSGENa}
\begin{split}
T = \frac{3}{4\pi z_h} - \frac{\mu^2 z_h}{16 \pi} - \frac{m^{2N} z_h^{2N-1}}{8\pi}   \,.
\end{split}
\end{align}

\paragraph{Translational symmetry breaking in holography:}
The holographic model \eqref{GENACa} has been used to study various translational symmetry breaking patterns: explicitly broken translational invariance (EXB), spontaneously broken translational invariance (SSB).

Solving the equations of motion near the AdS boundary ($z\rightarrow0$), one can check that the power of the potential in \eqref{GENACa}, $N$, determines the boundary behavior of the axion field\footnote{See also \cite{Jeong:2021zhz} for the description of the translational symmetry breaking with more general holographic models.}
\begin{align}\label{}
\begin{split}
\varphi_i \,=\, \varphi_i^{(0)} \,+\, \varphi_i^{(1)} \, z^{5-2N} \,+\, \dots \,,
\end{split}
\end{align}
where $\varphi_i^{(0)} = m \, x^i$ corresponds to the bulk solution \eqref{BSASC}.
According to the holographic dictionary, the leading term can be interpreted as the source and the sub-leading term is for the vacuum expectation value (vev).
Thus, when $N<5/2$, $\varphi_i^{(0)}$ is the source so that the translational symmetry is broken explicitly, via $\varphi_i^{(0)} = m \, x^i$. On the other hand, when $N>5/2$, $\varphi_i^{(0)}$ is no longer the source, instead it is the vev so one can study the spontaneously broken translational symmetry by taking $\varphi_i^{(1)} = 0$ with $N>5/2$. In summary, using the action \eqref{GENACa}, one can study the broken translational symmetry in holography as 
\begin{align}\label{MATTEOMODEL}
\begin{cases}
N< 5/2 :     \quad   \text{Explicitly broken translational symmetry}  \qquad\quad\, (\text{EXB}) \,, \\
N> 5/2 :     \quad   \text{Spontaneously broken translational symmetry}    \quad\, (\text{SSB}) \,.
\end{cases}
\end{align}

\paragraph{The entanglement density with broken translational symmetry.}
In what follows, we study $\sigma/s$ \eqref{SCEED} with various values of $N$. For this purpose, we identify $g(z_{*} \, u)$ from \eqref{BGMa} as
\begin{align}\label{genNgfor}
\begin{split}
g(z_{*} \, u) &= 1 - \tilde{\mathbf{g}}_{3} \, \tilde{z}_{*}^3 \, u^3 + \frac{\tilde{\mu}^2}{4} \tilde{z}_{*}^4 \, u^4  + \frac{\tilde{m}^{2N}}{2(2N-3)}  \tilde{z}_{*}^{2N}  u^{2N} \,,
\end{split}
\end{align}
where we used $\tilde{\mathbf{g}}_{3} := \mathbf{g}_{3} \, z_{h}^3$, $\tilde{\mu}:=\mu \, z_{h}$, and $\tilde{m}:=m \, z_{h}$.
Moreover, a similar relation as in \eqref{EDERS348} can also be used at finite $m$ 
\begin{equation}\label{TSBTMUBE}
\begin{split}
 \ell \, \mu \,=\, \tilde{\ell} \, \tilde{\mu}   \,, \qquad  \frac{T}{\mu} \,=\, \frac{1}{4\pi \tilde{\mu}}\left[ 3 - \frac{\tilde{\mu}^2}{4} - \frac{\tilde{\mu}^{2N}}{2} \left(\frac{m}{\mu}\right)^{2N} \right] \,,
\end{split}
\end{equation}
where $T$ is from \eqref{THMERRESULTSGENa}. Solving the relations \eqref{TSBTMUBE} one can find the expression of ($\tilde{\ell}, \tilde{\mu}$) in terms of ($\ell\,\mu, \, T/\mu, \, m/\mu, \, N$). Thus, $\sigma/s$ in \eqref{SCEED} can be evaluated as a function of  ($\ell\,\mu, \, T/\mu$) at given ($m/\mu, \, N$).

\subsection{First law of entanglement entropy}
Considering the $\tilde{z}_{*}\rightarrow0$ limit, let us first discuss FLEE \eqref{FLEENOUD} for general $N$. For easy comparison with the normal phase ($m=0$), we follow similar analysis as given in \eqref{smallLex}-\eqref{STE2DS}.

From the leading term of \eqref{genNgfor}, $g(z_{*} \, u)=1$, one can find the expression for small $\tilde{\ell}$ as \eqref{smallLex}.
Next, as in \eqref{smallC}, we also need to express $C(\tilde{z}_{*})$ by considering $g(z_{*} \, u)$ up to its sub-leading order. However, as can be seen from \eqref{genNgfor}, the sub-leading term of $g(z_{*} \, u)$ depends on $N$ as 
\begin{align}\label{smallCN}
\begin{cases}
N<3/2: \quad g(z_{*} \, u) &= 1 \,+\, \frac{\tilde{m}^{2N}}{2(2N-3)}  \tilde{z}_{*}^{2N}  u^{2N} \,+\, \dots \,, \\
N>3/2: \quad g(z_{*} \, u) &= 1 \,-\, \tilde{\mathbf{g}}_{3} \, \tilde{z}_{*}^3 \, u^3  \,+\, \dots \,, 
\end{cases}
\end{align}
where $\dots$ denotes higher order terms.\footnote{Note that the model \eqref{GENACa} may not be well defined at $N=3/2$, e.g., see \eqref{BGMa2}.}
For the $N>3/2$ case, we have the same sub-leading term of the normal phase ($m=0$) so that FLEE is obeyed at $N>3/2$ as \eqref{STE2DS}.

\paragraph{The violation of FLEE at $N<3/2$:}
However, for the $N<3/2$ case, we need to compute $C(\tilde{z}_{*})$ \eqref{HEEFORFIN2} with the different sub-leading term, $g(z_{*} \, u) = 1 \,+\, \frac{\tilde{m}^{2N}}{2(2N-3)}  \tilde{z}_{*}^{2N}  u^{2N}$, which produces
\begin{equation}\label{smallLexNV}
\begin{split}
C(\tilde{z}_{*}) &= -\frac{2\sqrt{\pi}}{3}\frac{\Gamma\left( \frac{7}{4} \right)}{\Gamma\left( \frac{5}{4} \right)} \,+\, \frac{\sqrt{\pi}}{32(3-2N)}\frac{\Gamma\left(\frac{N}{2}-\frac{1}{4}\right)}{\Gamma\left(\frac{N}{2}+\frac{5}{4}\right)} \, \tilde{m}^{2N}\,\tilde{z}_{*}^{2N} \,.
\end{split}
\end{equation}
Then, plugging \eqref{smallLexNV} into \eqref{STE1DS}, $\sigma/s$ can be further expressed as  
\begin{equation}\label{sigmaosN11}
\begin{split}
\frac{\sigma}{s} &\,=\,  \frac{3}{32(3-2N)}  \, \frac{\Gamma\left(\frac{5}{4}\right)\,\Gamma\left(\frac{N}{2}-\frac{1}{4}\right)}{\Gamma\left(\frac{7}{4}\right)\,\Gamma\left(\frac{N}{2}+\frac{5}{4}\right)} \, \left(\frac{m}{\mu}\right)^{2N} \tilde{\mu}^{2N} \, \tilde{z}_{*}^{2N-2}    \,,  \\
&\,=\,  \underbrace{  \frac{3}{32(3-2N)}  \, \frac{\Gamma\left(\frac{5}{4}\right)\,\Gamma\left(\frac{N}{2}-\frac{1}{4}\right)}{\Gamma\left(\frac{7}{4}\right)\,\Gamma\left(\frac{N}{2}+\frac{5}{4}\right)} \, \left( \frac{2\sqrt{\pi} \, \Gamma\left( \frac{3}{4} \right) }{ \Gamma\left( \frac{1}{4} \right) } \right)^{2-2N} \, \left(\frac{m}{\mu}\right)^{2N} \tilde{\mu}^{2}  }_{\,:=\, \mathcal{C}} \,\, (\ell\,\mu)^{2N-2} \,,
\end{split}
\end{equation}
where the sub-leading correction in \eqref{smallLexNV} only survives for the same reason explained above \eqref{STE2DS}.
In \eqref{sigmaosN11}, we also used $\tilde{m}=\frac{m}{\mu}\,\tilde{\mu}$ in the first equality and \eqref{smallLex}, \eqref{TSBTMUBE} in the last equality.

One can also find the higher order corrections to \eqref{sigmaosN11} by considering $g(z_{*} \, u)$ up to the $\mathcal{O}\left(\tilde{z}_{*}^3\right)$ order, $g(z_{*} \, u) = 1 \,+\, \frac{\tilde{m}^{2N}}{2(2N-3)}  \tilde{z}_{*}^{2N}  u^{2N} \,-\, \tilde{\mathbf{g}}_{3} \, \tilde{z}_{*}^3 \, u^3$, which gives \eqref{STE2DS} as a correction, i.e., 
\begin{equation}\label{sigmaosN22}
\begin{split}
\frac{\sigma}{s} \,=\,  \mathcal{C} \, (\ell \, \mu)^{2N-2}  \,+\, \left( \frac{\Gamma\left( \frac{1}{4} \right)}{\Gamma\left(\frac{3}{4}\right)} \right)^2 \, \frac{\tilde{\mathbf{g}}_{3}}{32 \, \tilde{\mu}} \, (\ell\,\mu)  \,+\, \dots\,,   
\end{split}
\end{equation}
where the leading term is \eqref{sigmaosN11}.
Therefore, for $N<3/2$, one can notice that FLEE is violated due to the leading term \eqref{sigmaosN11}: $\sigma/s \sim (\ell\,\mu)^{2N-2\,<\,1}$.
Note that when $N=1$ our result \eqref{sigmaosN22} reproduces \cite{Gushterov:2017vnr} in which the leading term \eqref{sigmaosN11} becomes a constant $\mathcal{C}$.

Based on the analysis \eqref{smallCN}-\eqref{sigmaosN22}, we find the violation of FLEE depending on the symmetry breaking pattern \eqref{MATTEOMODEL} as 
\begin{align}\label{RESULTOUR}
\begin{split}
\text{EXB}&:
\begin{cases}
\text{FLEE is violated} \,\,\,\quad (N<{3}/{2}) \,, \\  
\text{FLEE is obeyed} \,\,\,\,\,\, \quad ({3}/{2}<N<{5}/{2}) \,,  
\end{cases}
\\ \qquad
\text{SSB}&:
\,\,\,\,\,\text{FLEE is obeyed} \quad\quad (N>{5}/{2}) \,.
\end{split}
\end{align}
Note that our result \eqref{RESULTOUR} is complementary to the previous work \cite{Gushterov:2017vnr} ($N=1$ case), and shows that EXB does not guarantee the violation of FLEE, e.g., $3/2<N<5/2$.

\begin{figure}[]
 \centering
      \subfigure[$\sigma/s$ vs $\ell \, \mu$ \,at\, $N=1$ (EXB)]
     {\includegraphics[width=7.3cm]{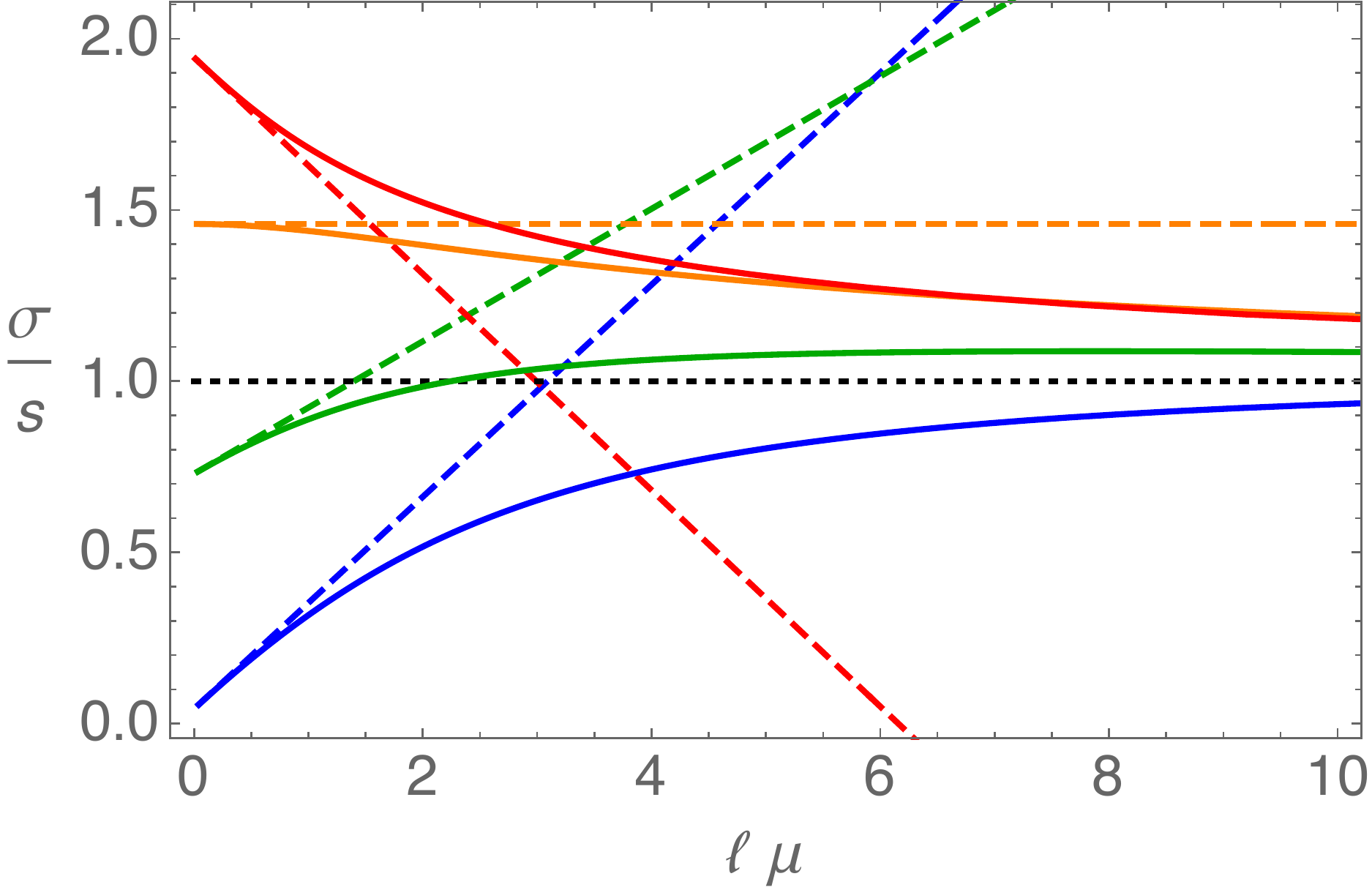}\label{EXB1a}} 
      \subfigure[$\sigma/s$ vs $\ell \, \mu$ \,at\, $N=3$ (SSB)]
     {\includegraphics[width=7.3cm]{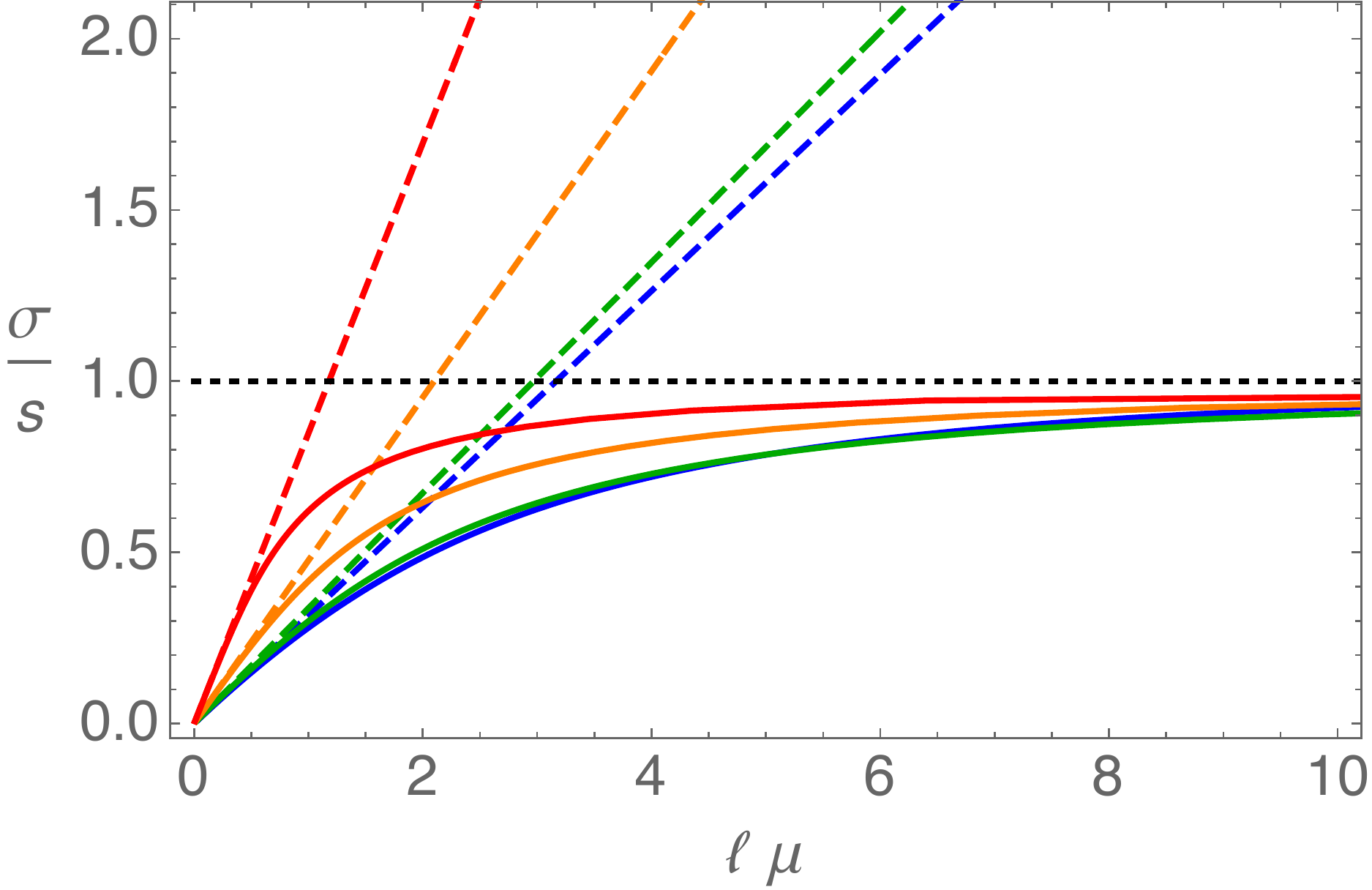} \label{SSB1b}} 
          \caption{Entanglement density at $T/\mu=0$, $m/\mu=(2, \,1, \,0.5, \,0)$ (red, orange, green, blue). The black dotted line is the guide line for $\sigma/s=1$. Solid lines are numerical results $\sigma/s$ via \eqref{SCEED}, dashed lines are analytic results: \eqref{sigmaosN22} in Fig. \ref{EXB1a}, \eqref{STE2DS} in Fig. \ref{SSB1b}. Note that the $m/\mu=0$ case corresponds to the blue data in Fig. \ref{FIGRN1}.} \label{EXBSSBFIG}
\end{figure}

In Fig. \ref{EXBSSBFIG}, we make the representative plots of $\sigma/s$ for FLEE with different symmetry breaking patterns,  $N=1$ (EXB) in Fig. \ref{EXB1a} and $N=3$ (SSB) in Fig. \ref{SSB1b},
in order to show that our analytic result is consistent with the numerical result of $\sigma/s$.\footnote{Note that our analytic analysis is valid for all $T$.}

\subsection{Area theorem}
Next, let us study $C(1)$ to check the area theorem \eqref{ARNOUD}. Note that, unlike the analysis for FLEE, we should resort to numerics in order for checking the area theorem~\cite{Gushterov:2017vnr}.

\begin{figure}[]
 \centering
      \subfigure[$C(1)$ vs $m/\mu$ at $N=(3, \,2.3, \,2.1, \,1.6)$ (red, \,orange, \,dashed black, \,green)]
     {\includegraphics[width=7.6cm]{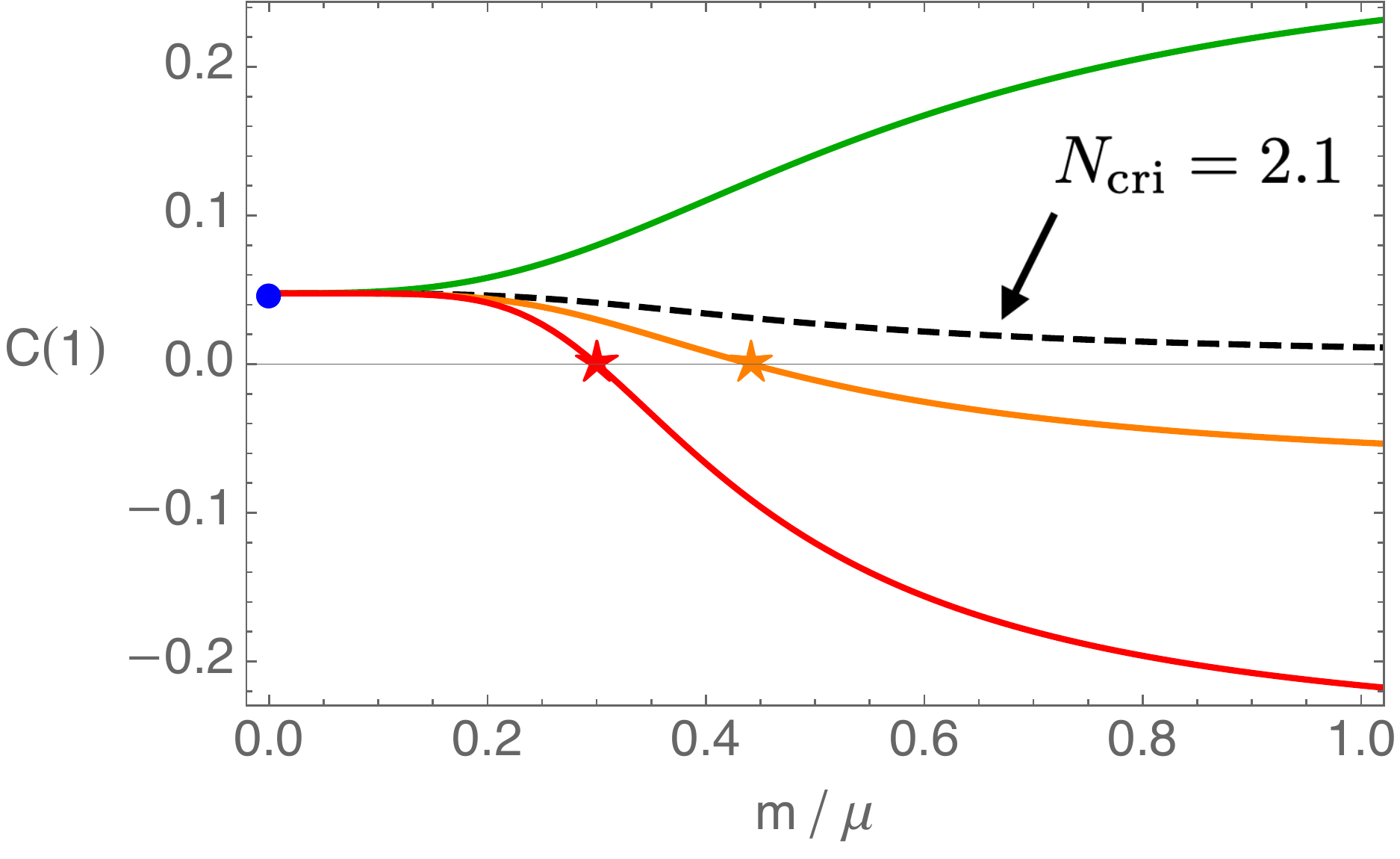}\label{EXBSSBFIG11}} 
      \subfigure[$m_{c}/\mu$ vs $N$]
     {\includegraphics[width=7.2cm]{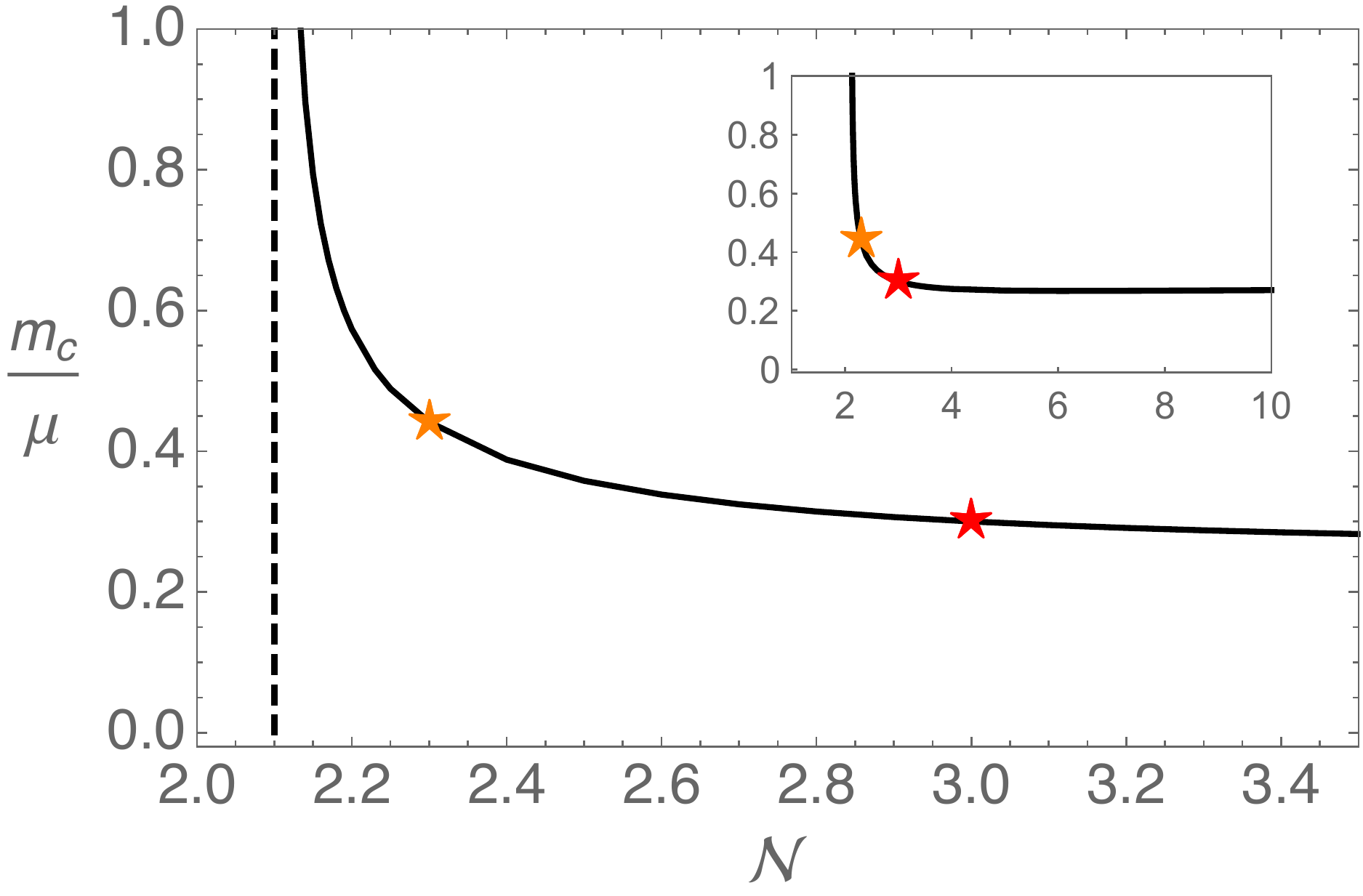} \label{EXBSSBFIG22}} 
          \caption{Checking the area theorem at $T=0$ with $m/\mu$. Blue dot in Fig. \ref{EXBSSBFIG11} is the same as in Fig. \ref{FIGRN2}. The stars represent $m_{c}/\mu$ and the black dashed lines are the data at $N_{\text{cri}}$.} \label{EXBSSBFIG33}
\end{figure}

In particular, we examine the area theorem by the sign of $C(1)$ at $T=0$ as in the normal phase ($m/\mu=0$) in Fig. \ref{FIGRN2}.\footnote{In all holographic models in this paper, we checked that $C(1)$ has the monotonic behavior with respect to $T$ (e.g., Fig. \ref{FIGRN}-\ref{scFIGSC}). We also checked that  $C(1)$ at finite $m/\mu$ has a monotonic behavior for $T$ as well, which is similar to Fig. \ref{FIGRN}. Thus the $T=0$ analysis of $C(1)$ may be enough to check the violation of the area theorem.} In other words, we study how the blue dot in in Fig. \ref{FIGRN2} behaves when we increase $m/\mu$ at given $N$. See Fig. \ref{EXBSSBFIG33}.

In Fig. \ref{EXBSSBFIG11}, we found that the translational symmetry breaking, a finite $m/\mu$, can change the sign of $C(1)$.
For instance, at $N=3$ (the red line in Fig. \ref{EXBSSBFIG11}), one can see that the positive value (blue dot) of $C(1)$ can become negative as we increase $m/\mu$. 
Here we denote the critical value of $m/\mu$ giving $C(1)=0$ as $m_{c}/\mu$: i.e., the area theorem is violated ($C(1)>0$) at $m<m_{c}$.\footnote{In Fig. \ref{EXBSSBFIG11}, the red/orange star represents $m_{c}/\mu$.}

From Fig. \ref{EXBSSBFIG11}, one can also find that $m_{c}/\mu$ (stars) depends on the value of $N$, i.e., $m_{c}=m_{c}(N)$. In particular, $m_{c}/\mu$ tends to increase as we decrease $N$, e.g., from $N=3$ (red star) to $N=2.3$ (orange star). As we decrease the value of $N$ further, one can find the critical $N$, $N_{\text{cri}}$, at which $C(1)$ is always non-negative (black dashed line). Here we numerically found $N_{\text{cri}}\sim2.1$. This $N$-dependent behavior of $m_{c}/\mu$ can be seen clearly in Fig. \ref{EXBSSBFIG22}.

From all the figures in Fig. \ref{EXBSSBFIG33}, we find the condition for the violation of the area theorem, $C(1)>0$, depending on the symmetry breaking pattern \eqref{MATTEOMODEL} as
\begin{align}\label{RESULTOUR2}
\begin{split}
\text{EXB}&:
\begin{cases}
\text{$C(1)>0$ \,at\, any $m$} \qquad\,\,\qquad (N<N_{\text{cri}}) \,, \\  
\text{$C(1)>0$ \,at\, $m<m_{c}(N)$} \qquad (N_{\text{cri}}<N<{5}/{2}) \,,  
\end{cases}
\\
\text{SSB}&:
\begin{cases}
\text{$C(1)>0$ \,at\, $m<m_{c}(N)$} \qquad (5/2<N<5) \,, \\  
\text{$C(1)>0$ \,at\, $m<m_{c}$} \qquad\,\,\,\,\,\quad (N\gtrsim5) \,,  
\end{cases}
\end{split}
\end{align}
where $N_{\text{cri}}\sim2.1$.\footnote{One can easily check the first line of \eqref{RESULTOUR2} in Fig. \ref{EXBSSBFIG11}: at $N<N_{\text{cri}}$, i.e., from a dashed black towards a green, $C(1)>0$ at any $m/\mu$.}
Note that at $N>N_{\text{cri}}\sim2.1$, $m_{c}$ is a function of $N$ in general as can be seen in Fig. \ref{EXBSSBFIG22}. However, we found that $m_{c}$ can be an $N$-independent universal value, $m_{c}/\mu\sim0.27$, at $N\gtrsim5$, which corresponds to the last line in \eqref{RESULTOUR2}. See the inset of Fig. \ref{EXBSSBFIG22}.

Based on \eqref{RESULTOUR} together with \eqref{RESULTOUR2}, we construct the summary table of the obedience/violation of FLEE/area theorem in the presence of the translational symmetry breaking. See Table. \ref{SUM2}.
\begin{table}[]
\begin{tabular}{| >{\centering\arraybackslash}m{1.75in} | >{\centering\arraybackslash}m{1.83in} | >{\centering\arraybackslash}m{1.85in}  |}
\hline
   Translational symmetry   & FLEE   & Area theorem  \\ 
 \hline
 \hline
     \small{Explicit breaking \quad ($N<{3}/{2}$)}     &  \small{Violated}      &  \small{Violated at any $m$}       \\
 \hline
      \small{Explicit breaking \quad (${3}/{2}<N<N_{\text{cri}}$)}     &  \small{Obeyed}      &  \small{Violated at any $m$}       \\
 \hline
     \small{Explicit breaking \quad ($N_{\text{cri}}< N<{5}/{2}$)}     &  \small{Obeyed}      &  \small{Violated at $m<m_c(N)$}       \\
 \hline
  \hline
    \small{Spontaneous breaking \quad (${5}/{2}<N<5$)}      &  \small{Obeyed}     &   \small{Violated at $m<m_c(N)$}         \\
  \hline
    \small{Spontaneous breaking \quad ($N\gtrsim5$)}      &  \small{Obeyed}     &   \small{Violated at $m<m_c$}         \\
    \hline
\end{tabular}
\caption{FLEE/area theorem with broken translation symmetry. The violation of the area theorem is examined at $T=0$ and we find $N_{\text{cri}}\sim2.1$. }\label{SUM2}
\end{table}
As in the U(1) symmetry breaking in Table. \ref{SUM1}, we find that the obedience/violation of FLEE/area theorem may also be used to classify the phases in which  translational invariance is broken spontaneously.
We also found the universality at $N\gtrsim5$ with the $N$-independent $m_{c}$.
Note that, like $M^2$ in holographic superconductors, for the SSB, $N$ can also be closely related to the material properties such as the shear modulus, $G(N)$,~\cite{Baggioli:2021xuv}.

\paragraph{Further comments on the violation of the area theorem:}
It is argued \cite{Gushterov:2017vnr} that the violation of the area theorem at low $T$ may be related to the IR geometry of black holes. For instance, the black hole with the AdS$_2\times R^2$ IR geometry \cite{Gushterov:2017vnr} turns out to violate the area theorem, i.e., $C(1)>0$.
However, our result \eqref{RESULTOUR2} may be one counter example for this argument because one can find that the area theorem can be obeyed ($C(1)<0$) at $m>m_{c}$ for the black hole \eqref{GENACa} in which its IR geometry is  the AdS$_2\times R^2$~\cite{Jeong:2021zsv}.\footnote{See the appendix in \cite{Jeong:2021zsv}, showing that the action \eqref{GENACa} has the AdS$_2\times R^2$ IR geometry.}

\paragraph{Entanglement entropy and Goldstone modes in holography:}
As mentioned in the introduction, entanglement entropy may have an $N_G \log(\ell/\epsilon)$ contribution from the Goldstone mode~\cite{Metlitski:2011pr,Rademaker_2015,Kulchytskyy:2015yea} where $N_G$ is the number of Goldstone modes.\footnote{Entanglement entropy has been investigated for coplanar antiferromagnets with $SO(3)$ symmetry in \cite{Metlitski:2011pr,Rademaker_2015}, the spin-1/2 $XY$ model \cite{Kulchytskyy:2015yea}.}
Let us close this section with a discussion for the appearance of such a $\log$ contribution in holography.
 
For SSB  $(N > 5/2)$ \eqref{MATTEOMODEL}, it is shown that there can be (transverse/longitudinal) phonons, the Goldstone modes of the translational symmetry, in holography \cite{Baggioli:2021xuv}. Thus, it will be instructive to see if SSB in holography can also show the $\log$ term in the entanglement entropy. 

Note that the $\log$ term comes with the cutoff divergence, $\log\epsilon$, so that we may need to investigate the UV-divergence structure in \eqref{USINT}, which can be analysed by the expansion of the integrand near the AdS boundary ($z\rightarrow0$) as
\begin{align}\label{CFIEMD44}
\begin{split}
\int^{z_{*}}_{\epsilon} \dd z {\frac{1}{z^2 \sqrt{1- \frac{z^4}{z_{*}^4} \frac{h(z_{*})^{2}}{h(z)^{2}} }}} \sqrt{\frac{h(z)}{g(z)}} \,=\, \int^{z_{*}}_{\epsilon} \dd z \left[ \frac{1}{z^2} \,+\, \frac{h'(0)-g'(0)}{2z}  \,+\, \dots \right]\,.
\end{split}
\end{align}
One can see that the leading term in \eqref{CFIEMD44} produces the usual UV divergence term $1/\epsilon$ and there can be an additional $\log$ divergence from the sub-leading term as $\left(g'(0)-h'(0)\right) \log \epsilon$ which is our interest.

However, for SSB ($N>5/2$), it turns out such a $\log$ term in the holographic model \eqref{GENACa} is vanishing in that $g(z)$ does not include a linear order in $z$ \eqref{BGMa} with $h(z)=1$, so $g'(0)-h'(0)=0$.
Thus, based on this result, we speculate that the $\log$ contribution of the entanglement entropy from the Goldstone mode may not appear in the strongly correlated systems.

%
\section{Conclusions}\label{sec5label}

We have studied entanglement entropy with the \textit{spontaneously} broken symmetry in holography.
In particular, using entanglement density \eqref{SIGFORF} in the small and large subsystem region, we examine if the obedience/violation of FLEE/area theorem \eqref{SIGFORFfin} can classify the phases in which the $U(1)$ or translational symmetry is broken. 

This kind of classification has been done for explicit-symmetry-breaking cases in \cite{Gushterov:2017vnr} and we extend it to spontaneous-symmetry-breaking cases. We show that indeed the obedience/violation of FLEE/area theorem may characterize the phases where the $U(1)$ or the translational symmetry is broken spontaneously. For the summary see Table. \ref{SUM1} or Table. \ref{SUM2}, where $M^2$ and $N$ may be related to the physical properties of materials such as $T_c$ or the shear modulus respectively.

Furthermore, we find some universalities from the  classification.
First, FLEE is always obeyed with the spontaneous symmetry breaking regardless of the type of symmetry: $U(1)$ or translational.
Second, independent of translational symmetry breaking patterns (EXB or SSB), the area theorem is always violated when the translational symmetry is weakly broken (small $m/\mu$).
Third, for SSB, in particular $N\gtrsim5$, the area theorem may not be related to the material properties (i.e., independent of $N$).  

As a byproduct, we find that the violation of the area theorem may not be related to the IR geometry unlike the speculation in~\cite{Gushterov:2017vnr}. 
At first we speculated that the condition for the violation of the area theorem may have something to do with the symmetry breaking but we found it was not the case. Understanding the precise condition for the violation of the area theorem remains an open problem and requires further investigation.
We also argue that the $\log$ contribution of the entanglement entropy from the Goldstone mode may not appear in strongly correlated systems.

Although the obedience/violation of the FLEE/area theorem can classify phases further within the spontaneously symmetry broken phase, it may not detect the phase transition between symmetry broken/unbroken phases (e.g., the normal phase vs the superconducting phase with $M^2\geq0$) or the symmetry breaking patterns (e.g., EXB at $N_{\text{cri}}<N<5/2$ vs SSB at $5/2<N<5$).
 
Based on this work, it will be interesting to investigate other quantum information properties, in the presence of spontaneous symmetry breaking, such as mutual information, entanglement wedge cross section, complexity~\cite{wipYW3}, and the entanglement entropy of the Hawking radiation, i.e., the Page curve.\footnote{See \cite{Chen:2021lnq,Casini:2022rlv} for a recent review of quantum information in holography.} 
We leave this subject as future work and will address them in the near future.

\acknowledgments

We would like to thank  {Sang-Eon Bak, Wen-Bin Pan, Yuan-Tai Wang}  for valuable discussions and correspondence.  
This work was supported by the National Key R$\&$D Program of China (Grant No. 2018FYA0305800), Project 12035016 supported by National Natural Science Foundation of China, the Strategic Priority Research Program of Chinese Academy of Sciences, Grant No. XDB28000000, Basic Science Research Program through the National Research Foundation of Korea (NRF) funded by the Ministry of Science, ICT $\&$ Future Planning (NRF- 2021R1A2C1006791) and GIST Research Institute(GRI) grant funded by the GIST in 2022.


\bibliographystyle{JHEP}

\providecommand{\href}[2]{#2}\begingroup\raggedright\endgroup

\end{document}